\documentclass[fleqn,usenatbib]{mnras}

\usepackage{newtxtext,newtxmath}

\usepackage[T1]{fontenc}
\usepackage{ae,aecompl}
\usepackage[dvipsnames]{xcolor}
\usepackage[toc,page]{appendix}
\usepackage[symbol]{footmisc}

\usepackage{graphicx}  
\usepackage{amsmath}  
\usepackage{colortbl} 

\usepackage{subcaption}
\DeclareUnicodeCharacter{2212}{\textendash}

\DeclareRobustCommand{\VAN}[3]{#2}
\let\VANthebibliography\thebibliography
\def\thebibliography{\DeclareRobustCommand{\VAN}[3]{##3}\VANthebibliography}

\title[Microlensing towards M31]{Using white dwarf lensing to resolve accretion flows}

\author[S L. Newman, M. Middleton, A. McMaster.]{
Sophie L. Newman$^{1,2}$\thanks{E-mail: sophie.newman@port.ac.uk}, Matthew J. Middleton$^{2} \, \&$ Adam McMaster$^{2}$ 
\\ 
$^{1}$ Institute of Cosmology and Gravitation, University of Portsmouth, Burnaby Road, Portsmouth PO1 3FX, UK\\
$^{2}$ Department of Physics and Astronomy, University of Southampton, Highfield, Southampton SO17 1BJ, UK }

\date{Accepted XXX. Received YYY; in original form ZZZ}

\pubyear{2023}

\begin{document}
\label{firstpage}
\pagerange{\pageref{firstpage}--\pageref{lastpage}}
\maketitle

\begin{abstract}
Microlensing is one of the most powerful tools for probing the nature of dark halo objects and the sources they lens. As our nearest massive galaxy, M31 provides a rich source population with many potential lenses in its halo crossing our field of view at any one time. In this paper we explore the probability that X-ray sources in M31 will be lensed by white dwarfs in M31's halo. We find an expected lensing rate of 2.6/year within the mean archival {\it Swift} XRT field-of-view, and 6.3/year for the whole galaxy. For X-ray emitting sources harboring accreting neutron stars and black holes, we find that microlensing offers a unique opportunity to constrain the properties of the inner accretion flow. Our results demonstrate that it is feasible to recover both the spin of the black hole and the temperature profile of the accretion disk by discerning their effects upon the profile of the microlensing magnification. We show that these parameters have a significant effect on the shape of the light curve, with the effect of spin being more pronounced at smaller impact parameters and higher energies, while the effect of the temperature profile is larger at lower energies and larger impact parameters. This suggests that multi-band observations of a single lensing event could be used to robustly constrain both parameters.

\end{abstract}

\begin{keywords}
accretion -- accretion disks -- gravitational lensing: micro -- white dwarfs -- X-rays: binaries
\end{keywords}



\section{Introduction}



Gravitational lensing is the well-observed effect of deflected null geodesic paths from background sources by massive foreground objects that act as a lens. In the strong lensing regime, the mass density of the lens is high enough to cause multiple images, arcs, or rings to form. Observations of multiple images in galaxy-scale strong lenses have led to competitive constraints on cosmological parameters \citep{collett,holicow,hogg,tian}, studies into the nature of galactic and sub-galactic scale dark matter halos \citep{koopmans,gilman,nadler,shajib,ballard}, and developments into our understanding of the formation and evolution of massive galaxies \citep{mandelbaum,peng,leauthaund,sonnenfeld,chan,etherington}. 

In the case of microlensing, the foreground lens is less massive and is often a transiting compact object such as a neutron star (NS), black hole (BH), white dwarf (WD), star, or even exoplanet \citep{exoplanets}. In contrast to strong gravitational lensing, the separated and deformed images of the source cannot be resolved, and instead we observe the summed magnification of each of the images as the lens transits the source. The lensing event itself encodes a great deal of information about both the lens \citep{Gould2000,Wyrzykowski2020} and the source being lensed \citep[e.g. the prevalence of star spots;][]{sunspots}.



The most common targets for microlensing studies are the Large and Small Magellanic Clouds (LMC and SMC) and the Galactic bulge \citep[e.g.][]{udalski94,sumi13,mroz19,han24,nunota24} due to the large number of background stars.
Notable results include the recent discovery of the first isolated stellar-mass black hole using any technique \citep{isolatedBH,sahu25} and limits placed on the black hole mass distribution in the Milky Way (MW) halo \citep{blaineau}. The study by \cite{kruszynska24} found events using \textit{Gaia Data Release 3} \citep[\textit{DR3},][]{dr3} mainly towards the Galactic center and estimated the mass of the lenses using information obtained from their best-fitting microlensing models, finding eleven candidates for dark remnants. Likewise, \cite{2023Gaia} also used \textit{Gaia DR3} data to find 363 microlensing events, of which 90 had not been reported before, with the majority found towards the Galactic bulge. 


Beyond the LMC, the next nearest microlensing target is M31. It is an order of magnitude more distant than the LMC, which leads to sources being typically unresolved for ground-based telescopes. This has led to the requirement to study the lensing signal within individual pixels (pixel lensing), as first suggested by \cite{Gould1996}. Both the \textit{Microlensing Exploration of the Galaxy and Andromeda} (\textit{MEGA}) collaboration \citep{MEGA2006} and the \textit{POINT-AGAPE} survey \citep{POINT-AGAPE2003} have reported multiple lensing events towards M31 (14 and 4 candidate events respectively) through such campaigns.

Whilst the smaller projected size of M31 means that it should be transited by fewer Galactic halo objects \citep{Calchi2012}, there are three main advantages to targeting M31 for microlensing searches. Firstly, M31 is estimated to have a stellar mass roughly ten times that of the MW \citep{mwmass,m31mass} which means that it has a much larger population of potential lensing objects, such as stars and stellar remnants, in its halo. The higher density of these objects increases the overall probability of a microlensing event. Secondly, M31 is favorably oriented with respect to our line of sight, allowing us to view a large, unobscured portion of its stellar population, providing a rich field of potential sources for lensing. Combining these first two factors, \cite{Crotts1992} estimate there should be approximately 13 times more lensing events of sources in M31 by its own halo lenses than by lenses in the Milky Way's halo. Thirdly, the greater distance to M31 also results in very small lens proper motions relative to those located in the MW, resulting in long-duration events that are easier to monitor with typical all-sky surveys.  These factors make M31 an ideal target for microlensing surveys, despite its smaller angular size on the sky.




Since lensing is achromatic, it is not only optical sources which can be lensed. X-ray bright systems, such as X-ray binaries (XRBs) containing accreting neutron stars and black holes, will also be lensed and this could be detectable in X-ray observations. At the level of most microlensing events (a factor of a few in magnification) and typical instrumental sensitivities, we would expect the only X-ray bright systems in M31 detectable under lensing to be fairly well-isolated, circumventing one of the major issues with studying lensing of sources in that galaxy. This presents a significant advantage for microlensing surveys in dense stellar environments, such as M31, effectively circumventing the problem of source blending that complicates analysis in crowded optical fields.

Lensing of accreting compact objects is also a powerful route to access new physical insights, as it has the potential to resolve the otherwise unresolvable accretion flow. The use of this capability has been limited to lensing of quasars \citep{Dai2003,Pooley2007,Kochanek2007,Morgan2008,chartas,Dai2010} where the X-ray emission is found to emerge from within $\sim 10 R_{\rm g}$, and the optical emission at 4000\AA \, has been constrained to emerge from a region $\sim 70 R_{\rm g}$ around a $10^8 M_\odot$ BH. Until now, the potential of lensing of accreting binary systems has not been explored. We do this in a similar fashion to \cite{heyrovsky}, who used microlensing to determine the limb darkening of a lensed star, where variations in magnification as the source passes behind the lens reveal details about the source's surface brightness profile. Analogously, the flux from an accretion disk should peak at different radii depending on the band in which it is observed, and this should yield signatures in the microlensing signal.

In this paper we present the first estimate for the rate of X-ray microlensing by halo objects in M31, and explore the potential of lensing for constraining the otherwise unresolvable structure of the accretion disc around accreting compact objects.



\section{Methods}

Here we consider lensing only by white dwarfs as these objects far outnumber neutron stars and black holes and can have a high proper motion. For example, \cite{Fantin2021} find 90 MW halo WD candidates from the Canada-France Imaging Survey, by considering an object to be
part of the Galactic halo if it has a transverse velocity greater than 200 km/s. Additionally, \cite{torres19} identify 95 halo WD candidates based on a 100 pc sample from \textit{Gaia-DR2}, with a mean tangential velocity of 197 km/s and a maximum of 484 km/s. 

To estimate a reasonable number of WDs for our lensing study we draw on results from binary population synthesis (Section \ref{sec:popsynth}) and then proceed to estimate the chance probability of lensing of X-ray sources in M31, localised through deep observations taken by NASA's {\it Chandra} (Section \ref{sec:pointsourcepopulation}). For computational ease, we will initially consider a smaller sample of objects located in the MW halo and then scale to lenses in the halo of M31 (Section \ref{sec:sims}) using the well established differences in total stellar mass and distance. Finally in Section \ref{sec:accrectionflow}, we calculate band-limited flux maps of accretion flows and obtain the lensing signatures that may permit new insights into accretion using a variety of instruments and surveys.

\subsection{Population synthesis}
\label{sec:popsynth}

\begin{figure}
    \centering
    \includegraphics[width=\linewidth]{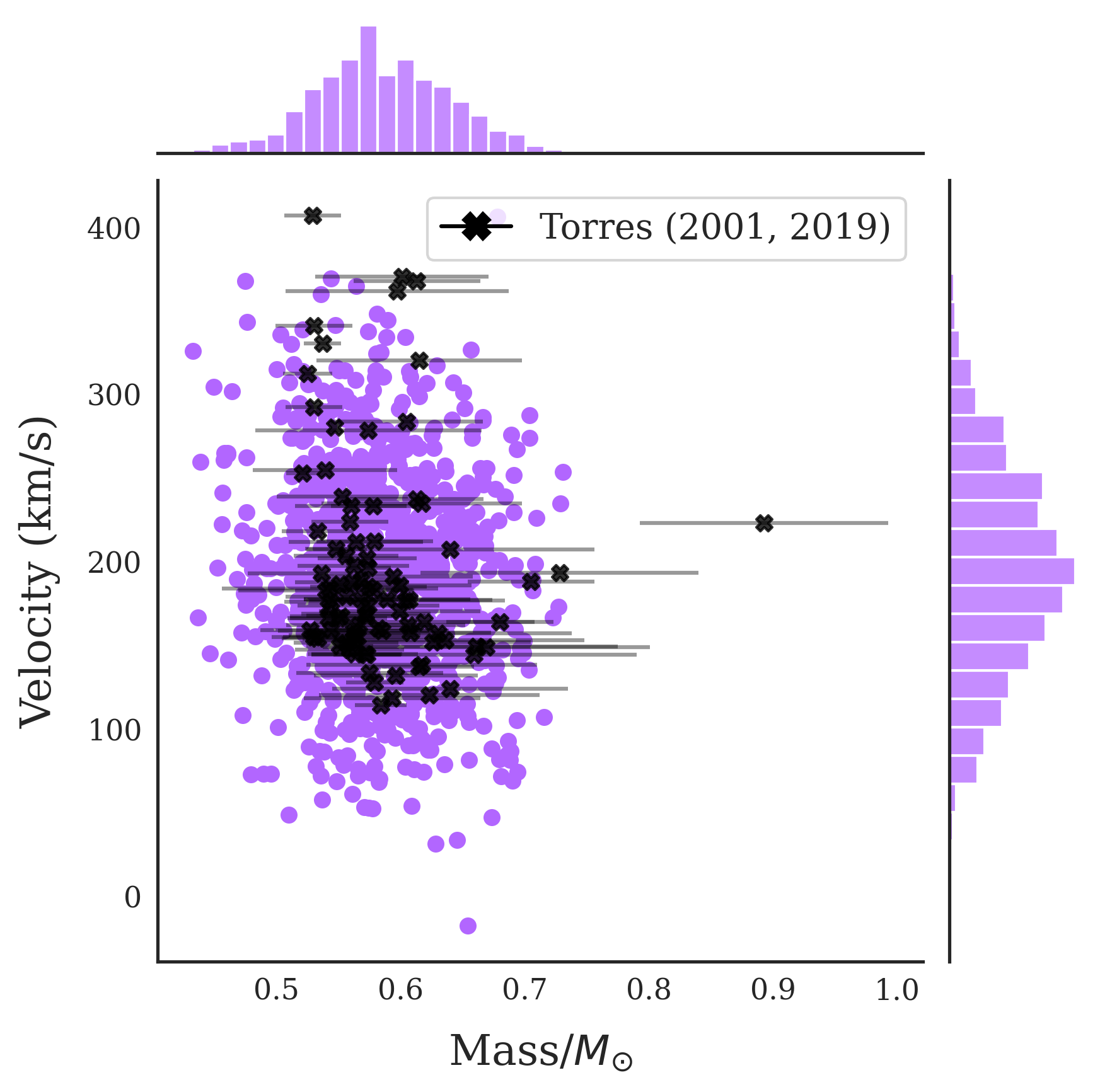}
    \caption{Sample MW halo WD masses and velocities drawn from a 2D Gaussian distribution, derived from data from \protect\cite{torres19} and \protect\cite{torres21}. Here dark blue crosses and corresponding error bars show the original data derived by Torres with photometric methods. Lighter lilac circles show points randomly sampled from the Gaussian distribution.}
    \label{fig:wd_dist}
\end{figure}



In this work we utilise the estimated numbers of WDs from \cite{ruiter} obtained using {\sc StarTrack} (\citealt{startrack}), a stellar population synthesis code that was created with a strong emphasis placed on calculating populations of compact objects. {\sc StarTrack} performs detailed calculations of several key astrophysical processes, such as stellar wind mass loss rates, the loss of angular momentum in binaries due to gravitational radiation, the decrease of rotation of a binary component due to magnetic braking, and the evolution of orbital parameters due to tidal interactions between binary components. {\sc StarTrack} has been used in a variety of studies, from modeling the gravitational-wave background created by double compact objects (\citealt{2021PhRvD.103d3002P}), to determining the numbers of self-lensing binaries within the MW (\citealt{wikselflensing}). 


The number of WD lenses in the halo of the MW towards M31 ($N_{\rm lenses,MW}$) can be crudely estimated using the projected area containing the sources in M31: 
\begin{align}
    N_{\rm lenses,MW} = \frac{N_{\rm total,MW}}{4\pi} \theta^2
    \label{eq:mw_lenses}
\end{align}

\noindent where $\theta$ is the field-of-view (FOV) of our chosen instrument. As we are focusing on the rate of X-ray microlensing, we assume the instrument properties for NASA's {\it Swift XRT}, i.e. $\theta$ = 15 arcmin \citep{watson09}. Once the total number of MW lenses is known (${N_{\rm total,MW}}$), the number within our FOV  can then be found. 

We assume that the MW and M31 have a similar IMF; \cite{IMF} studied radial gradients of optical and NIR IMF-sensitive features along the major axis of the bulge of M31,
out to a distance of $\approx$ 800 pc and found an implied IMF for the M31 bulge consistent with a MW-like distribution. Whilst it is unknown how the star formation history compares between the two galaxies, we will also assume them to be similar. Knowing that M31 is approximately ten times more massive than the MW (\citealt{mwmass}, \citealt{m31mass}), we assume that M31 will have ten times more lenses, and we therefore assume a simple scaling of: 
\begin{align}
    \frac{N_{\rm lenses,M31}}{N_{\rm lenses,MW}} = 10 \left(\frac{D_{\rm M31}}{D_{\rm MW}}\right)^2
    \label{eq:ratio_of_observed_lenses}
\end{align}


\noindent where $D_{M31}$ and $D_{MW}$ are the distances to the M31 and MW halo respectively. Estimates for $N_{\rm total,MW}$ range from $10^8$ by \cite{ruiter} (through use of {\sc StarTrack}) to the highly-debated value of over $10^{11}$ from a previous microlensing study towards the LMC by \cite{alcock}. 
Combining the more conservative value of $10^8$ by \cite{ruiter} with equations \ref{eq:mw_lenses} and \ref{eq:ratio_of_observed_lenses}, we obtain 127 WD lenses in the halo of the MW towards M31 in the FOV of {\it Swift} and $7.7 \times 10^6$ lenses in the halo of M31 within the same field. 


In addition to the number of WD lenses, we require their velocity and mass distributions. For these we refer to  \cite{torres19}, where a Random Forest algorithm was used to classify objects from \textit{Gaia-DR2} out to a distance of 100 pc. To train the algorithm before applying it to the \textit{Gaia} data set, they use their own population synthesis code (\citealt{garcia-berro}) to generate a WD population according to a Galactic model with three components (thin disc, thick disc, and halo). This led to the identification of 95 halo WD candidates within the observed sample, along with their proper motions. Following this work, \cite{torres21} built the WD luminosity and mass function for the same sample. We model the masses and velocities from \cite{torres21} with a Gaussian distribution (see Figure \ref{fig:wd_dist}) with a best-fit mean of $0.58 M_\odot$, consistent with the original mass distribution. We perform subsequent draws from this distribution and create mock populations of WD lenses in subsequent simulations.

\subsection{X-ray point source population of M31}
\label{sec:pointsourcepopulation}

\begin{figure}
    \centering
    \includegraphics[width=\linewidth]{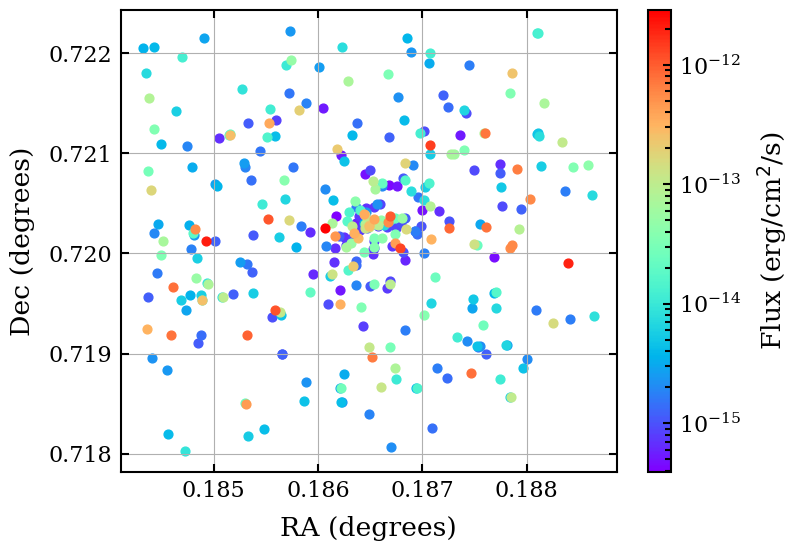}
    \caption{The flux distribution of X-ray sources in the stacked M31 observations by \protect\cite{catalogue} within the mean FOV of {\it Swift}. }
    \label{fig:m31_map}
\end{figure}

The time-dependent, X-ray point source population of M31 (detectable by current instruments) is composed of outbursting low mass X-ray binaries (LMXBs), persistent high mass X-ray binaries (HMXBs), cataclysmic variables, and recurrent novae. 
The deepest map of M31 at the time of writing was obtained by \cite{catalogue} who combined 133 \textit{Chandra ACIS-I/S} observations totalling $\approx 10^6$ seconds and detected 795 X-ray sources. The flux distribution of these 795 sources is shown in Figure \ref{fig:m31_map}. We note that, although the population and flux distribution is time-dependent, we will assume for the sake of ease that the 795 sources from \cite{catalogue} are persistent and will discuss the impact of the population demographic further in Section \ref{sec:discussion}.



\section{Lensing simulations for the X-ray population of M31}
\label{sec:sims}

\begin{figure}
    \centering
    \includegraphics[width=\linewidth]{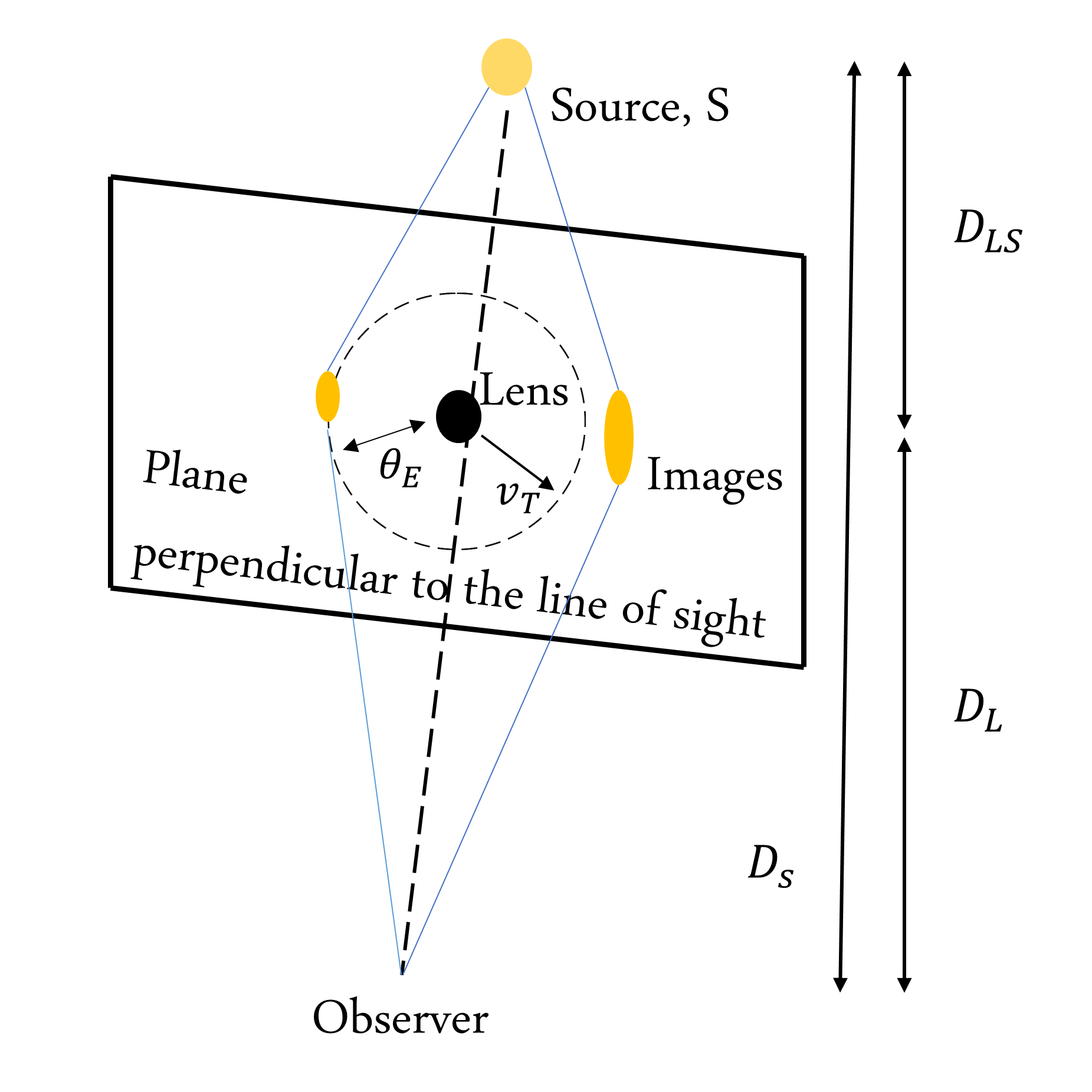}
    \caption{Geometric diagram for a lens moving at transverse velocity $v_{\rm t}$ across a source as seen in the plane perpendicular to the observer's line-of-sight.}
    \label{fig:distances}
\end{figure}

\begin{figure*}
    \centering
    \begin{subfigure}{0.45\linewidth}
        \includegraphics[width=\linewidth]{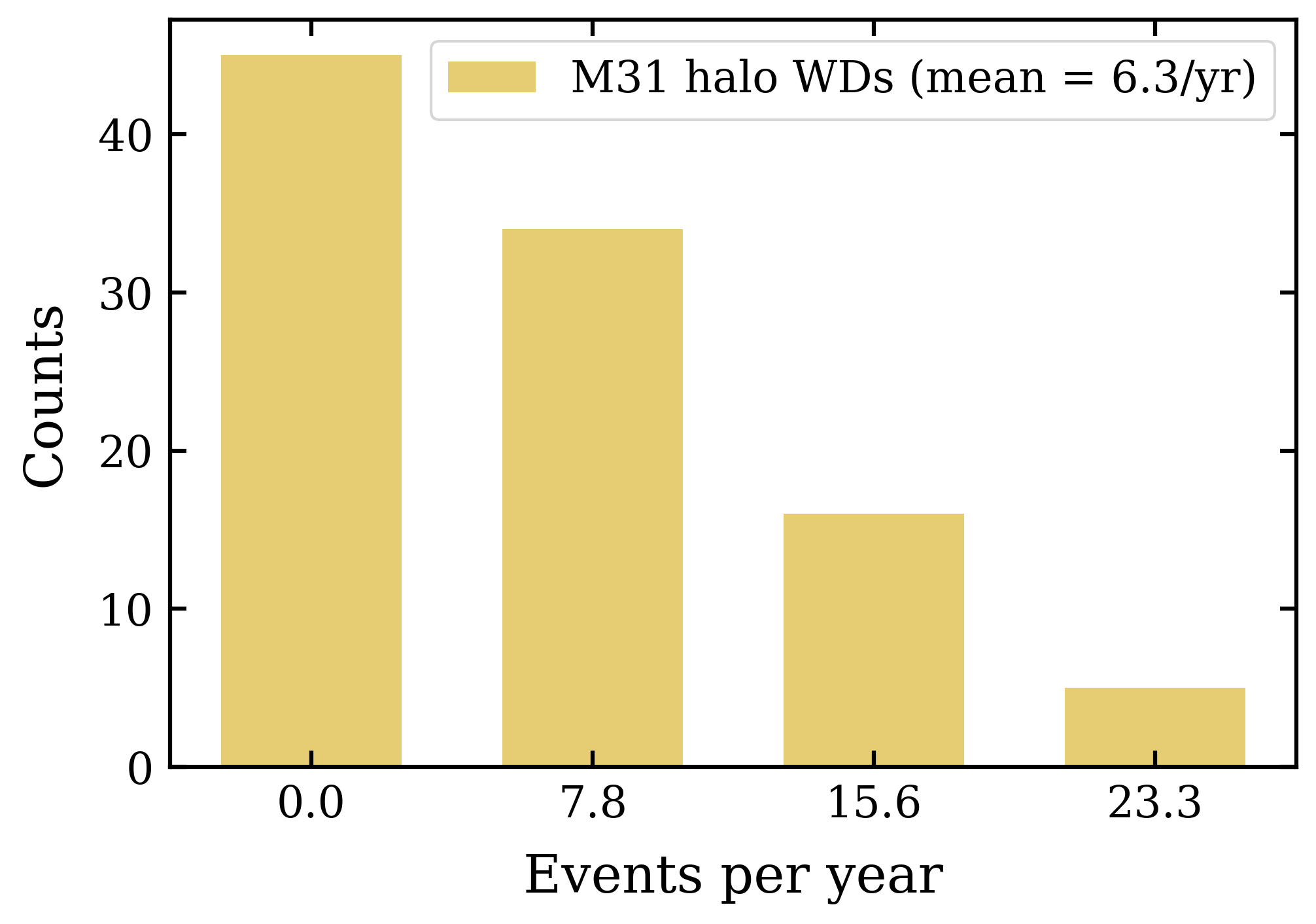}
        \caption{}
        \label{fig:event_dist}
    \end{subfigure}
    \hfill
    \begin{subfigure}{0.45\linewidth}
        \includegraphics[width=\linewidth]{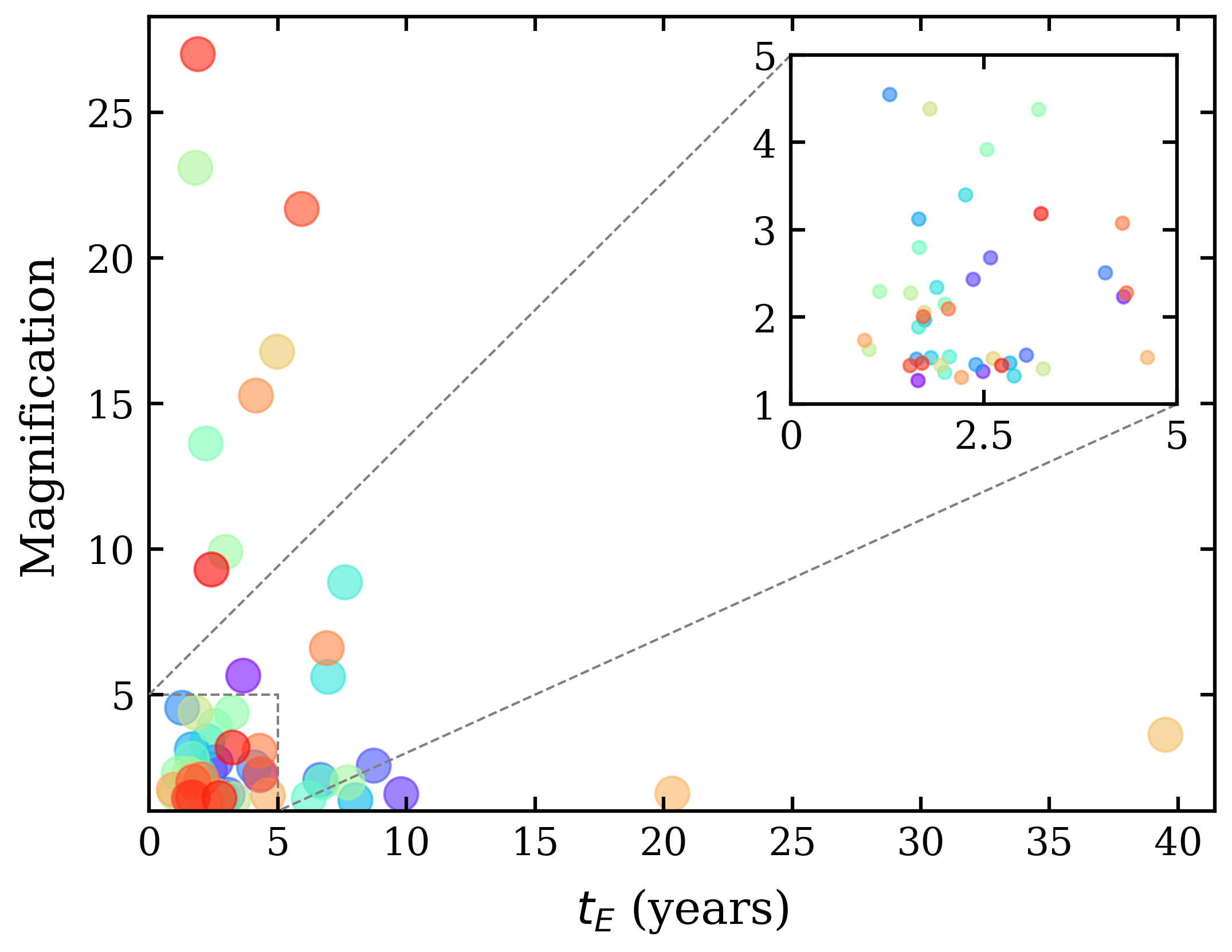}
        \caption{}
        \label{fig:magsvst}
    \end{subfigure}
    \caption{Figure a) shows the number of lensing events per year for a total lensing population of $7.7 \times 10^6$ WD lenses in the halo of the M31. Figure b) shows the magnification $A$ and $t_{\rm E}$ for each simulated event after the crossing times are scaled for a population of lenses in the halo of M31 using $t_{E} \propto \sqrt{d}$ with an M31 lens distance of 780 kpc and 10 kpc for MW lenses. The inset zoom highlights the higher density of events with $t_{\rm E} < 5$ yr and $A < 5$.}
    \label{fig:event_characteristics}
\end{figure*}

\begin{figure}
    \centering
    \includegraphics[width=\linewidth]{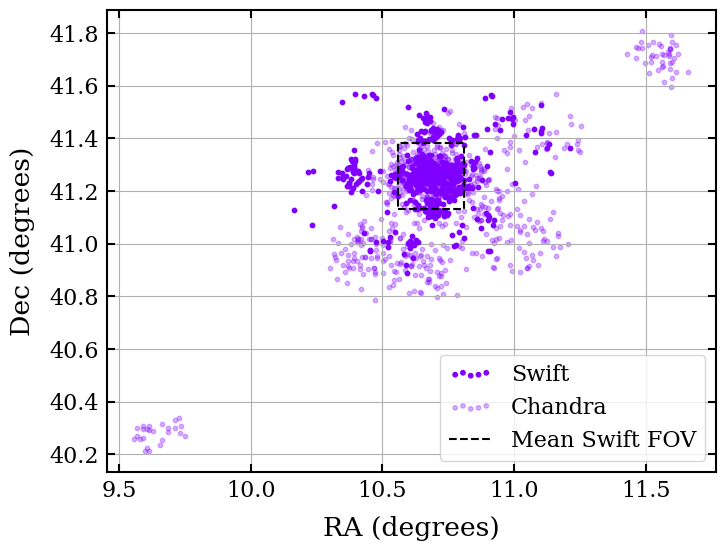}
    \caption{Map of the M31 \textit{Chandra} sources from the catalogue by \protect\cite{catalogue} and the {\it Swift} pointing positions from September 2006 to March 2023. All of the observations available at the time of writing are shown, and the {\it Swift} XRT FOV is overlaid around the mean {\it Swift} XRT pointing position.}
    \label{fig:map}
\end{figure}

We simulated $3 \times 10^8$ WD lenses in the halo of the MW, moving across the projected region in M31 harbouring the X-ray sources. The masses and space velocities of the lenses were sampled from a Gaussian distribution (see Figure \ref{fig:wd_dist}). To simplify computations, we used 20\% of the background sources in M31 and later scaled the results linearly based on the desired number of sources. We chose to simulate MW lenses, rather than those in the halo of M31, because their greater proper motions reduce the computational time. These results were then re-scaled for M31 lenses. We simulated a significantly larger number of MW lenses than expected to permit resampling.

The azimuthal angle $\phi$ and the polar angle $\theta$ of the WD's velocity vector in the plane perpendicular to the line-of-sight (as shown in Figure \ref{fig:distances}) are drawn from a uniform distribution. The transverse velocity $v_{\rm t}$ of a given lens is then found in the standard manner:
\begin{align*}
    v_{\rm x} &= v \sin(\theta)\cos(\phi) \\
    v_{\rm y} &= v \sin(\theta)\sin(\phi) \\
    v_{\rm t} &= \sqrt{v_{\rm x}^2 + v_{\rm y}^2} 
\end{align*}
\noindent where $v$ is the space velocity of the lens (e.g. from a kick or dynamical interaction). The transverse velocity of the lens is then used to calculate the proper motion:
\begin{align}
    \mu = v_{\rm t}/d
    \label{eq:proper_motion}
\end{align}

\noindent where $d$ is the observer distance to the lens, which, for distances between 10 and 100 kpc gives proper motions for MW WD lenses $\sim 10^{-3} - 10^{-4}$ arcsec/yr.


For our simulations, we assume that halo lenses have a density which decreases with distance from the Galactic center as:
\begin{align}
    \rho \propto \left(1 + \frac{r}{a_{\rm 0,halo}}\right)^{-3.5}
    \label{eq:density}
\end{align}
\noindent where $a_{\rm 0,halo}$ is the scale radius of 3.5 kpc \citep{Zinn,morrison}. A starting position for a given lens on the sky is randomly selected and its trajectory is then calculated using the proper motion calculated in Equation \ref{eq:proper_motion}. 

Following standard formulae, a source at a distance $D_{\rm S} = 780$ kpc and a lens at a distance $D_{\rm L}$ (drawn between 10 and 100 kpc according to its number density in Equation \ref{eq:density}) results in a total amplification of the source (i.e. when the two images are not resolvable) of:
\begin{align}
    A = \frac{b(t)^2 + 2}{b(t)\sqrt{b(t)^2 +4}}
    \label{eq:A}
\end{align}
\noindent where $b(t)$ is the angular separation between the lens and source in units of the angular Einstein radius given by: 

\begin{align}
   \theta_E = \sqrt{\frac{4GM(D_S-D_L)}{c^2D_SD_L}} 
   \label{eq:einstein_angle}
\end{align}
\noindent where $M$ is the lens mass. For WD lenses in the MW halo, this gives $\theta_{\rm E}$ $\sim 10^{-4}$ arcsec, whereas M31 lenses have $\theta_{\rm E} \sim 10^{-5}-10^{-6}$ arcsec.

As described in \cite{nutshell}, the variation of magnification with time due to the lens moving across the source is determined by the impact parameter, b(t):
\begin{align}
    b(t) = \sqrt{b^2_0+\left(\frac{(t-t_0)}{t_E}\right)^2}
\end{align}
where $b$ is the projected distance of closest approach of the lens to the source, $t_0$ is the time at which the lens crosses the source, and $t_{\rm E}$ is the characteristic Einstein crossing time of the event, $t_{\rm E} = \theta_{\rm E}/\mu$. For each point in a given lens' trajectory, we determine whether it lies within $\theta_{\rm E}$ of the source; if so we obtain $b$, the magnification, and $t_{\rm E}$. 

\subsection{Lensing the entire X-ray population of M31}

In the above we have assumed the 795 sources identified by \cite{catalogue} to be our background sources for potential lensing. From our simulations, we find 64 events occur within this larger FOV for the $3 \times 10^8$ lenses simulated. We then resample for a population of $3 \times 10^6$ lenses in the halo of the MW and find a mean of 0.81 events over 16 years, as illustrated in Figure \ref{fig:event_dist}. 

To estimate the number of events resulting from lenses in the halo of M31, the following simple scaling was used: 

\begin{align}
    \frac{N_{\rm events,M31}}{N_{\rm events,MW}} \approx \frac{N_{\rm lenses,M31}}{N_{\rm lenses,MW}} \frac{\theta_{\rm E,M31}}{\theta_{\rm E,MW}} \frac{\mu_{\rm E,M31}}{\mu_{\rm E,MW}} 
\end{align}

After rescaling the results to the expected population of $7.7 \times 10^6$ M31 WD halo lenses, we predict an event rate of 6.3 events/year. 
After rescaling the crossing times to those expected by M31 lenses (using $t_{E} \propto \sqrt{d}$ with an M31 lens distance of 780 kpc and 10 kpc for MW lenses), the resulting distribution of $t_{\rm E}$ and peak magnification is shown in Figure \ref{fig:magsvst}.


\subsection{Archival X-ray microlensing events with {\it Swift}}
\label{sec:sw}

\begin{figure*}
    \centering
    \begin{subfigure}{0.45\linewidth}
        \hspace{-1cm} 
        \includegraphics[width=\linewidth]{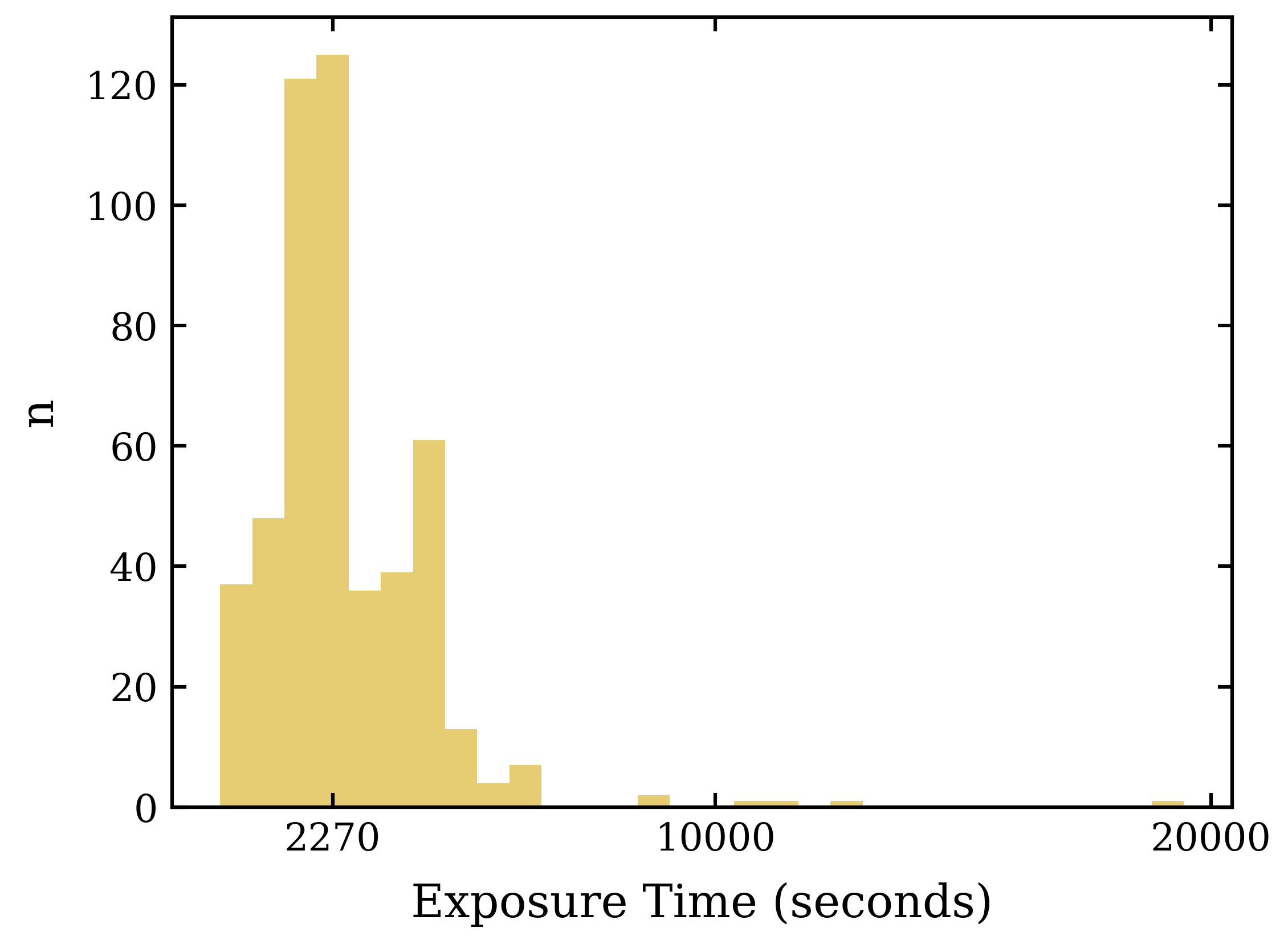}
        \caption{}
        \label{fig:exposuretimes}
    \end{subfigure}
    \begin{subfigure}{0.43\linewidth}
        \hspace{0.5cm} 
        \includegraphics[width=\linewidth]{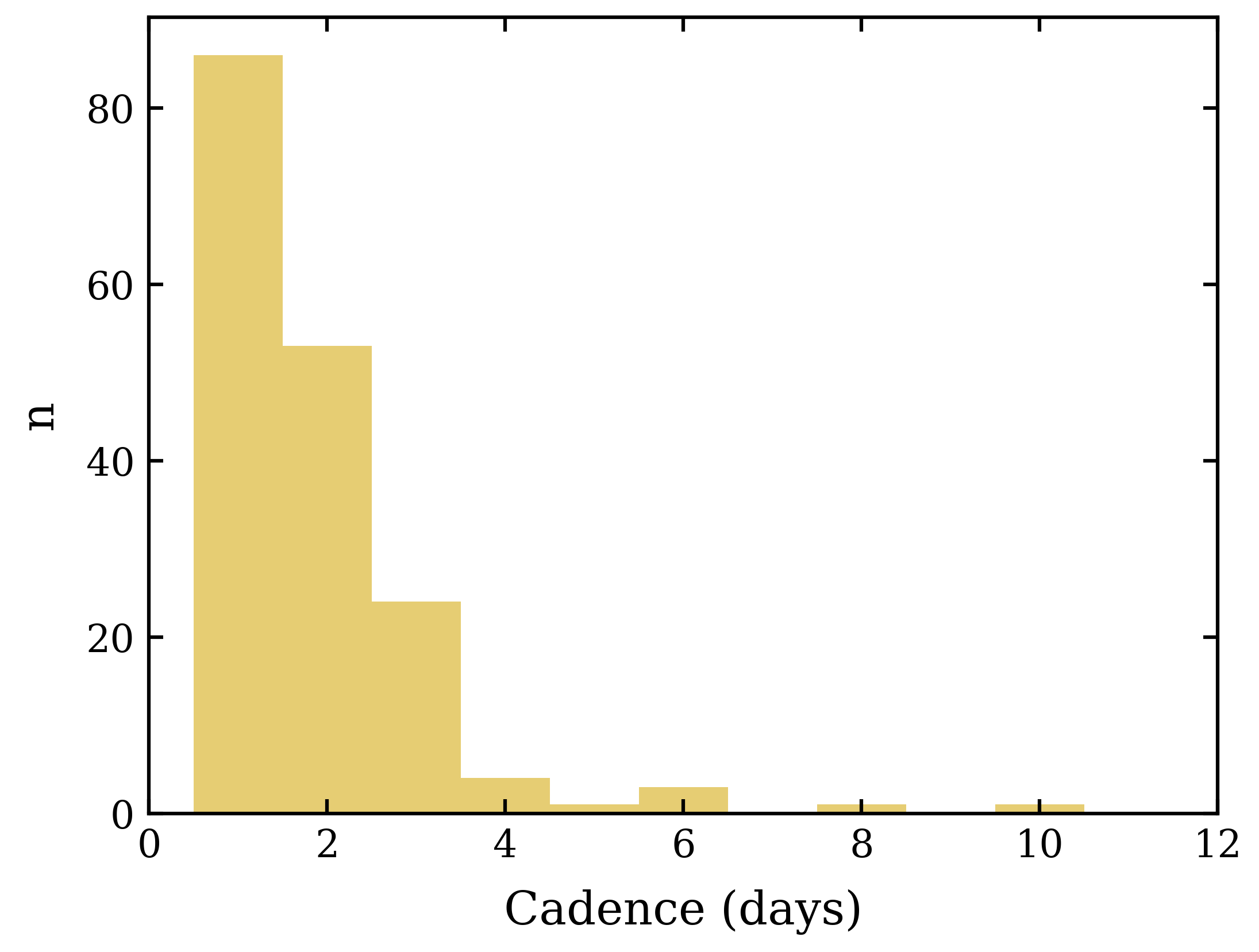}
        \caption{}
        \label{fig:cadence}
    \end{subfigure}
    
    \caption{Characteristics of the {\it Swift} observations towards M31 from September 2006 to March 2023 (all of the observations available at the time of writing), within the mean {\it Swift} pointing position shown in Figure \ref{fig:map}. The modal values for exposure time and cadence are 2270s and 1.1 days respectively. }
    \label{fig:swift_data}
\end{figure*}

In the above, we made predictions for lensing of the entire detectable X-ray population of M31. At the time of writing there are a remarkable 16 years of {\it Swift} data available in the archive (a total of 740 observations at the time of writing), representing a long baseline in which to search for X-ray microlensing (which will be the focus of a forthcoming paper). In order to explore the potential of this data, we require the number of {\it Chandra} sources expected within a {\it typical} {\it Swift} observation towards M31, which we obtain by finding the mean pointing position. In Figure \ref{fig:map} we plot the {\it {\it Swift}} pointing positions along with the {\it Chandra} source positions (\citealt{catalogue}), onto which is projected the {\it Swift} XRT FOV. We observe that $67\%$ of the total number of {\it Swift} pointing positions and 42\% of the \textit{Chandra} sources lie within the average $15 \times 15$ arcmin FOV. 
In Figure \ref{fig:swift_data}, the distribution of exposure times and cadences of the {\it Swift} observations in this average field are plotted, with modal values of 2270s and 1.1 days respectively. The flux distribution of the 334 \textit{Chandra} sources making up the assumed possible lensed sources are shown in Figure \ref{fig:m31_map} (and we stress once again that we have assumed there is no time-dependence). From our simulations above, 
we would infer that, given the 334 sources in the mean {\it Swift} FOV, we should expect effectively zero events/year due to lenses in the MW halo and a mean of 2.6 events/year due to lenses in the halo of M31.

For each simulated lensing event in the previous section, a randomly chosen flux for a source lying within the mean {\it Swift} FOV (taken from the catalogue of \citealt{catalogue}) is assigned as the event's initial flux (as any of the sources may be lensed without preference) before the simulated magnification is applied. We subsequently determine whether a given event is detectable by {\it Swift} by assuming a detection sensitivity of $8.8 \times 10^{-14}$ erg/cm$^2$/s for the modal {\it Swift} exposure time (\citealt{burrows}). 
Using this flux limit and resampling each event, we find that around 31\% of events are above {\it Swift}'s detection sensitivity. This indicates that there should be around 0.8 events/year observable by {\it Swift} in the mean FOV, equivalent to 12.7 events for the 16 years of {\it Swift} data available towards M31.







\begin{figure*}
    \centering
    \begin{subfigure}{0.45\linewidth}
        \includegraphics[width=\linewidth]{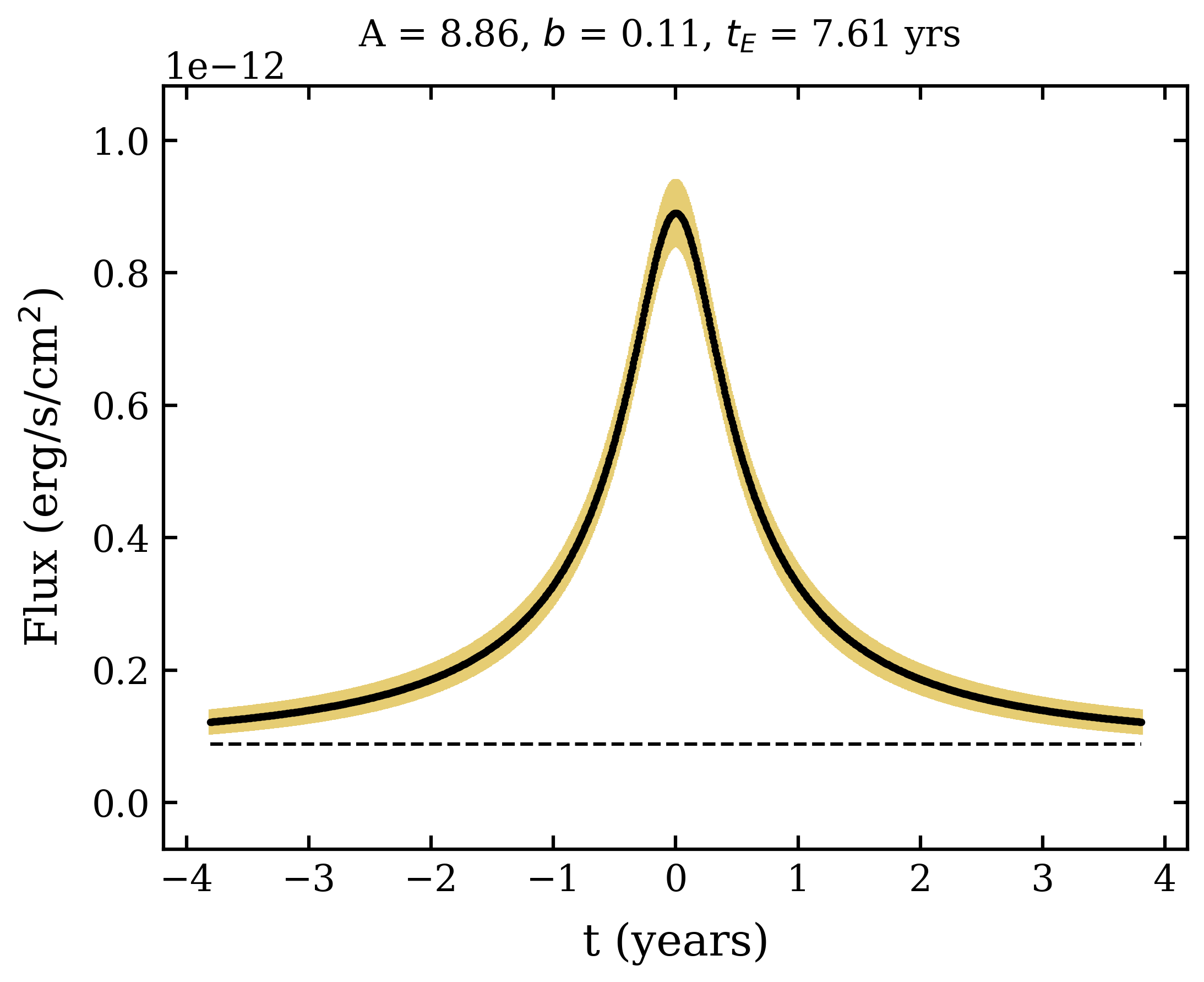}
        \caption{}
        \label{fig:events1}
    \end{subfigure}
    \hfill
    \begin{subfigure}{0.45\linewidth}
        \includegraphics[width=\linewidth]{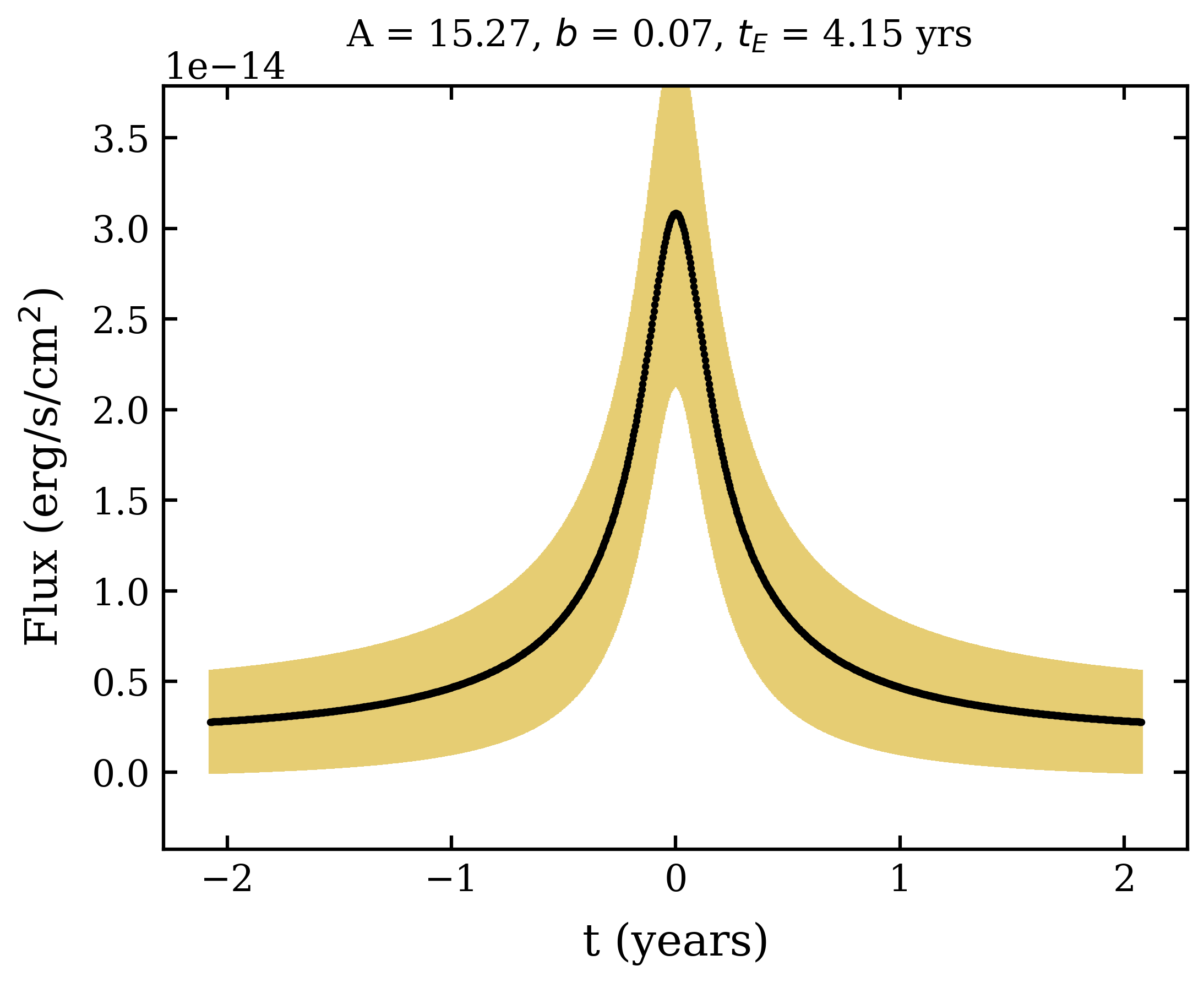}
        \caption{}
        \label{fig:events2}
    \end{subfigure}
    
    \vspace{\baselineskip}
    
    \begin{subfigure}{0.45\linewidth}
        \includegraphics[width=\linewidth]{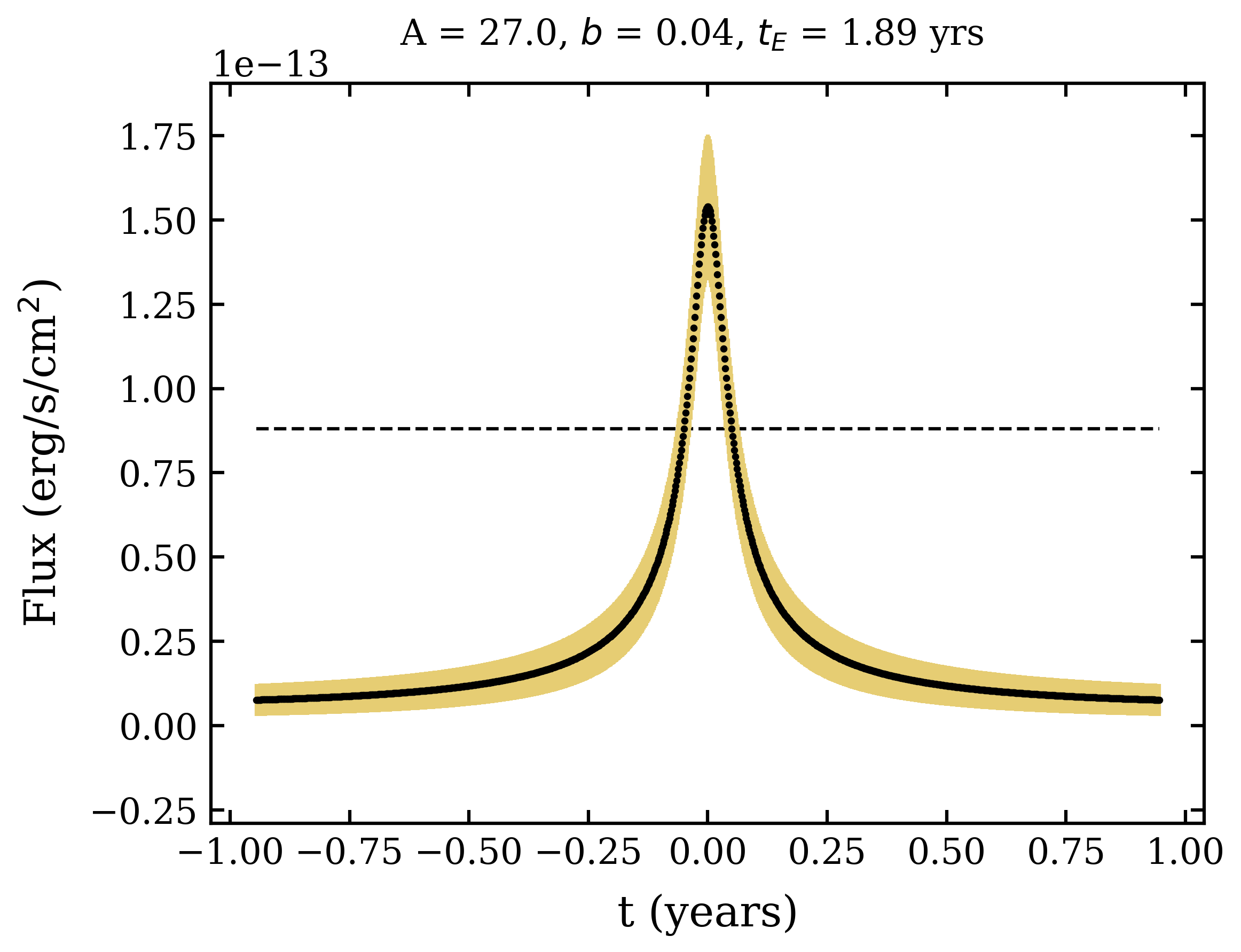}
        \caption{}
        \label{fig:events3}
    \end{subfigure}
    \hfill
    \begin{subfigure}{0.45\linewidth}
        \includegraphics[width=\linewidth]{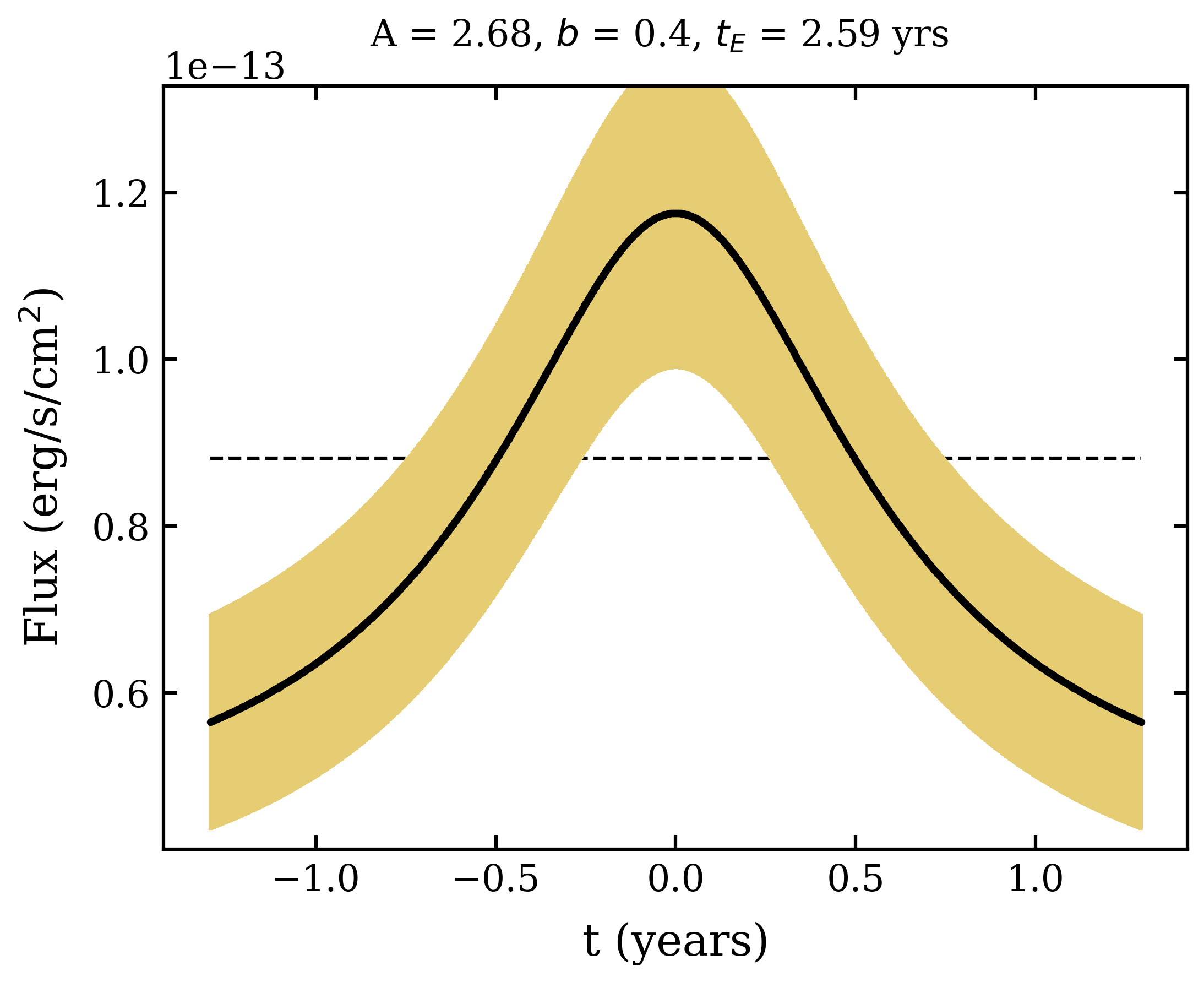}
        \caption{}
        \label{fig:events4}
    \end{subfigure}

    \caption{A selection of simulated microlensing events by WD lenses in the halo of M31. The dashed black line indicates {\it Swift}'s sensitivity for an average exposure time of 2270 seconds. 1 sigma errors, based on the inferred count rates, are shown. Event characteristics are provided in each case with the maximum magnification of the event, $A$, impact parameter $b$, and crossing time $t_{E}$.}
    \label{fig:events}
\end{figure*}

Examples of events -- both above and below the {\it Swift} sensitivity threshold -- are shown in Figure \ref{fig:events}, assuming the average {\it Swift} cadence and exposure time. In Figures \ref{fig:events3} and \ref{fig:events4} we show how an event's magnification can lead to a source, not usually observable by {\it Swift} (but observable by \textit{Chandra}), being detected due to lensing.


\section{Resolving the accretion flow}
\label{sec:accrectionflow}

\begin{figure*}
    \centering
    \begin{subfigure}{0.45\linewidth}
        \hspace{0.005cm} 
        \includegraphics[width=\linewidth]{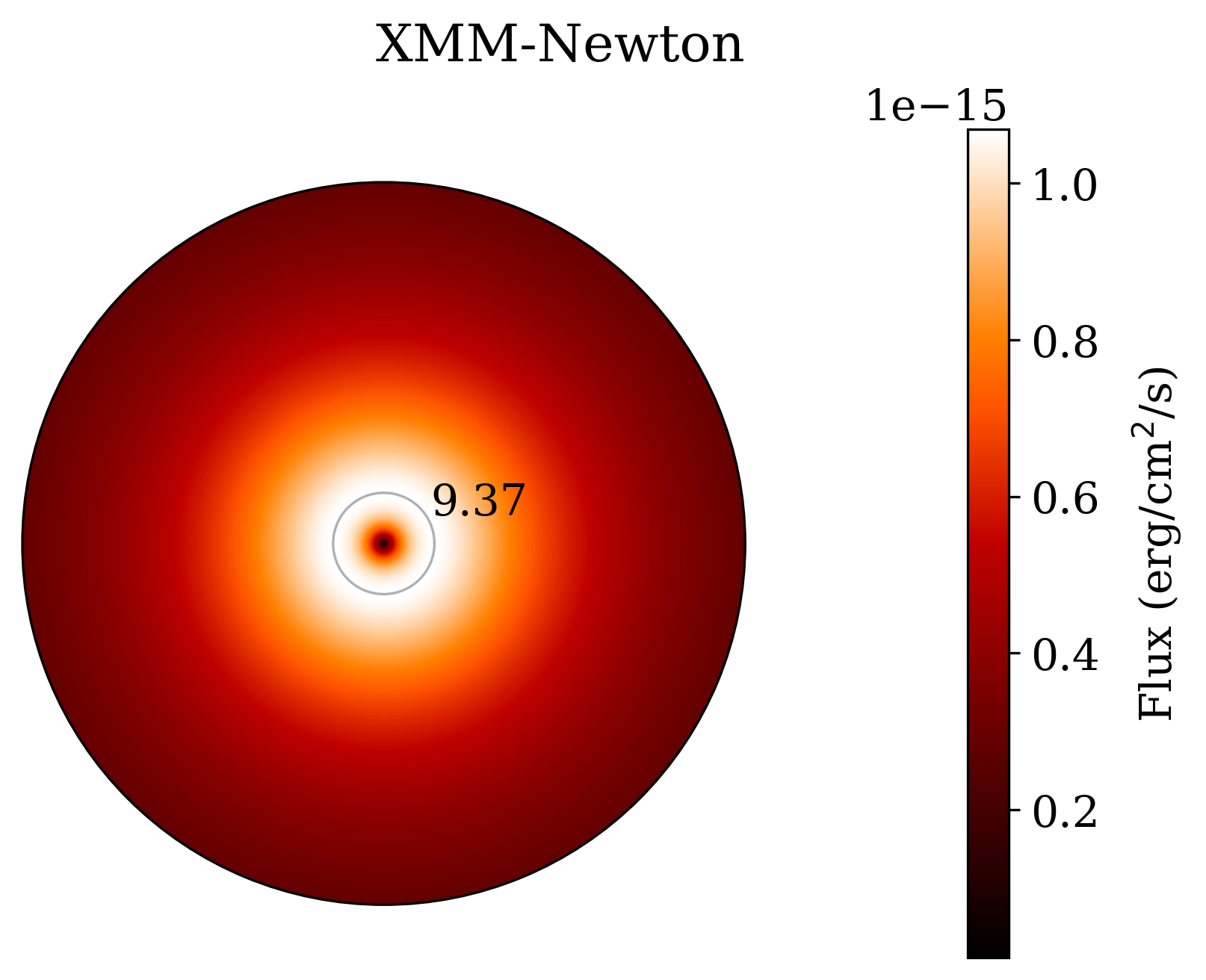}
        \caption{}
        \label{fig:xmm}
    \end{subfigure}
    \hfill
    \begin{subfigure}{0.45\linewidth}
        \includegraphics[width=\linewidth]{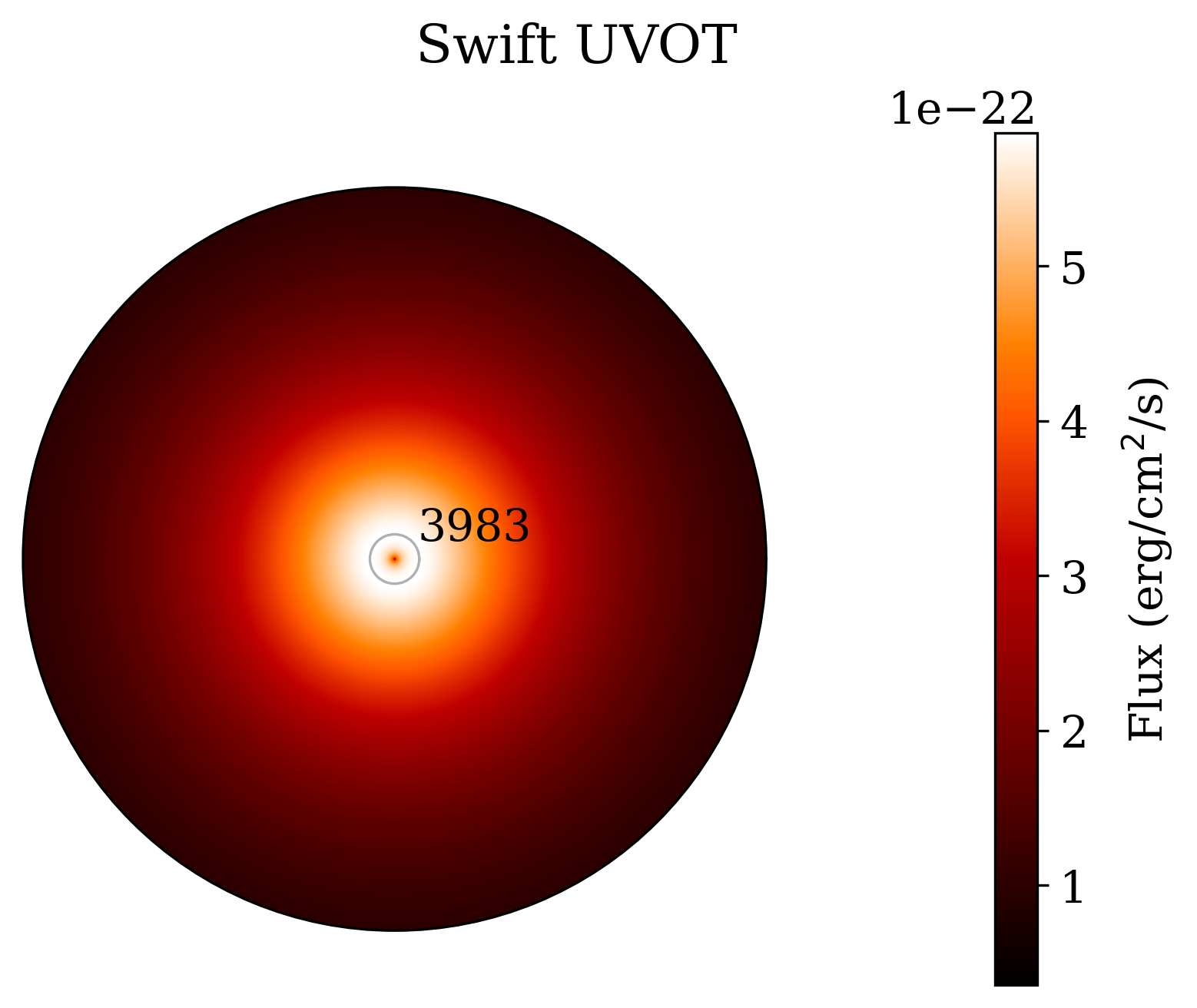}
        \caption{}
        \label{fig:uvot}
    \end{subfigure}
    
    \vspace{\baselineskip}
    
    \begin{subfigure}{0.45\linewidth}
        \includegraphics[width=\linewidth]{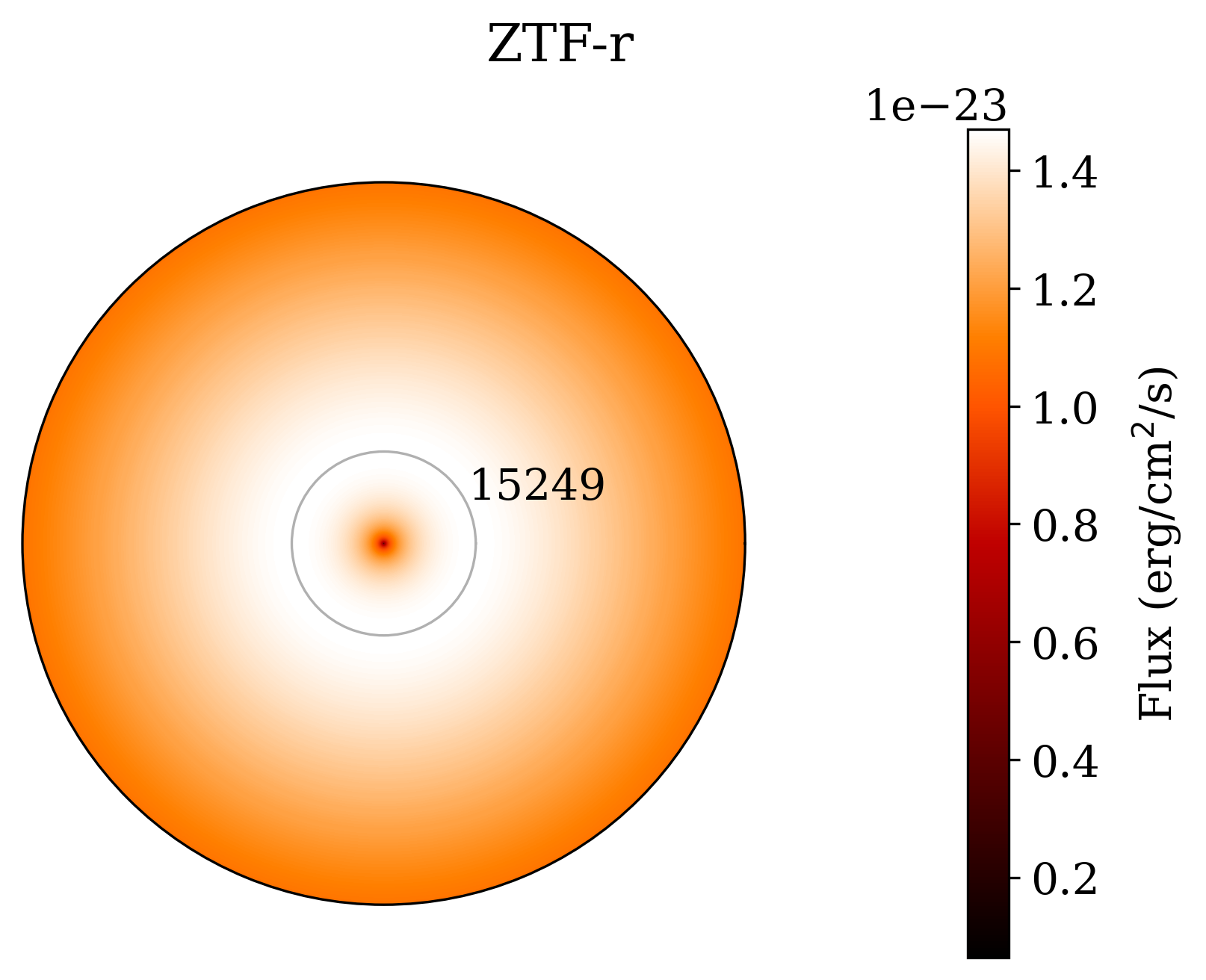}
        \caption{}
        \label{fig:ztf_r}
    \end{subfigure}
    \hfill
    \begin{subfigure}{0.45\linewidth}
        \includegraphics[width=\linewidth]{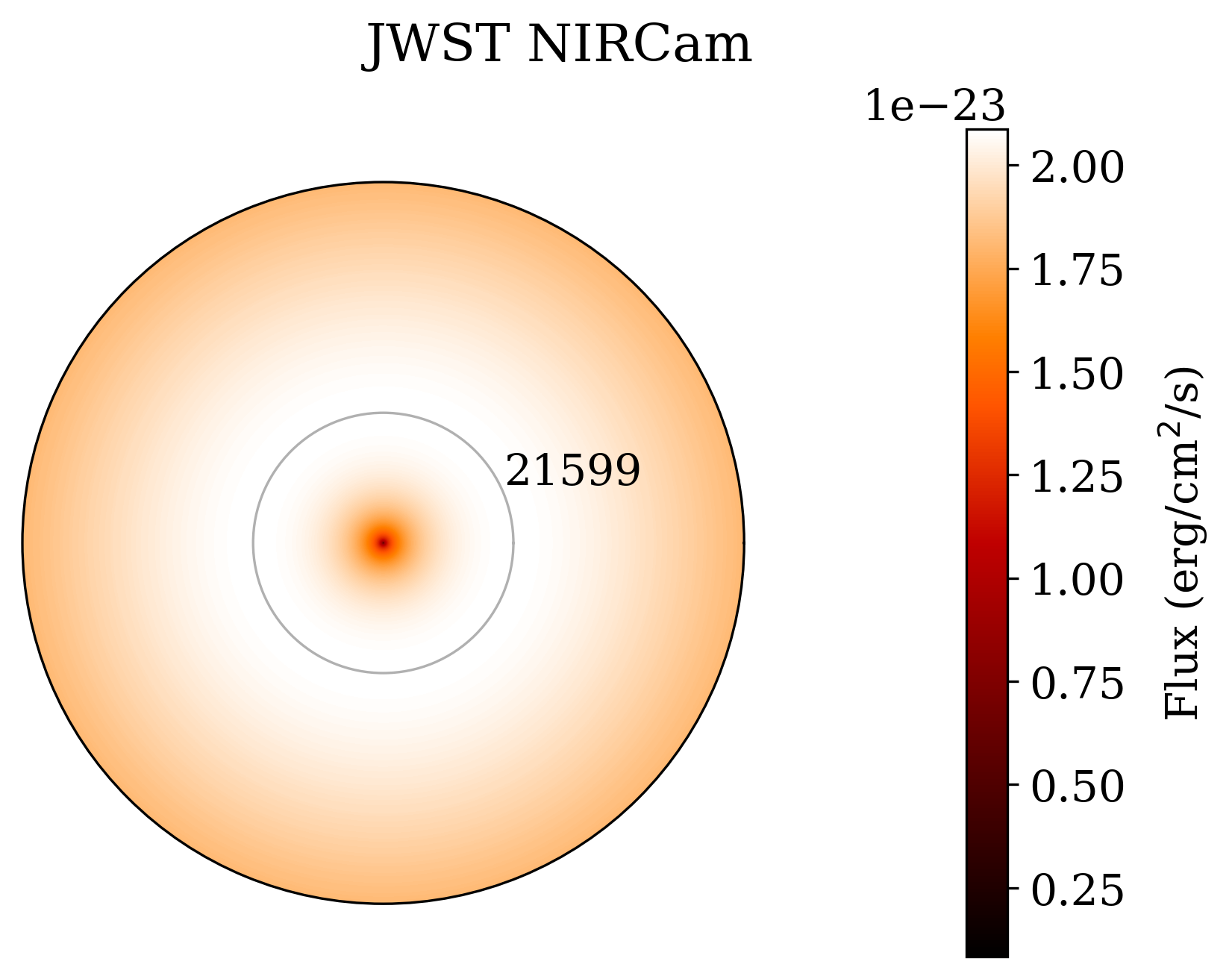}
        \caption{}
        \label{fig:jwst}
    \end{subfigure}
    
    \caption{Flux maps of a geometrically thin, optically thick accretion disc around a $10 M_{\odot}$ black hole with zero spin ($a = 0$) and a temperature coefficient of $\alpha = 3$. Maps are presented for wavelength bands accessible to \textit{JWST}, \textit{ZTF}, {\it Swift UVOT}, and {\it XMM-Newton}. The innermost radius plotted is $1.25 R_{\rm g}$ and the outermost radius plotted for each is $\approx 10^{6} R_{\rm g}$, except for {\it XMM-Newton} which is $\approx 10 R_{\rm g}$ due to the disk's compactness. The thin, blue circle shows the radius of maximum flux in each case in units of $R_{\rm g}$. Here we see that {\it XMM-Newton} observes the brightest emission within a small disk, while \textit{JWST} has fainter emission from a more extended disk.}
    \label{fig:heatmap}
\end{figure*}

\begin{figure*}
    \centering
    \includegraphics[width=\linewidth]{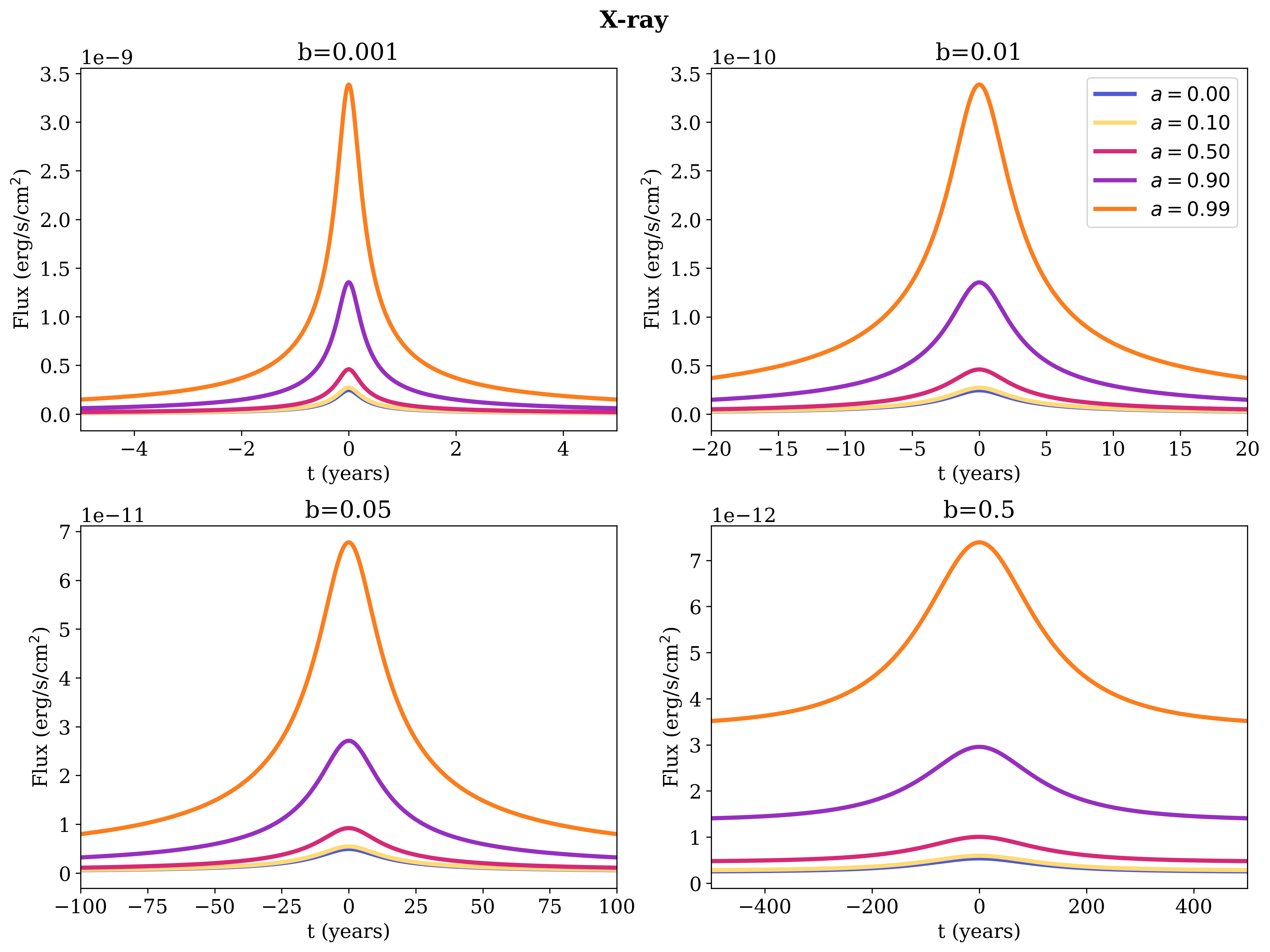}
    \caption{Simulated lensing events in the X-ray band, illustrating how impact parameter and spin  affect the magnification over time.}
    \label{fig:mag-xray-spin}
\end{figure*}

\begin{figure*}
    \centering
    \includegraphics[width=\linewidth]{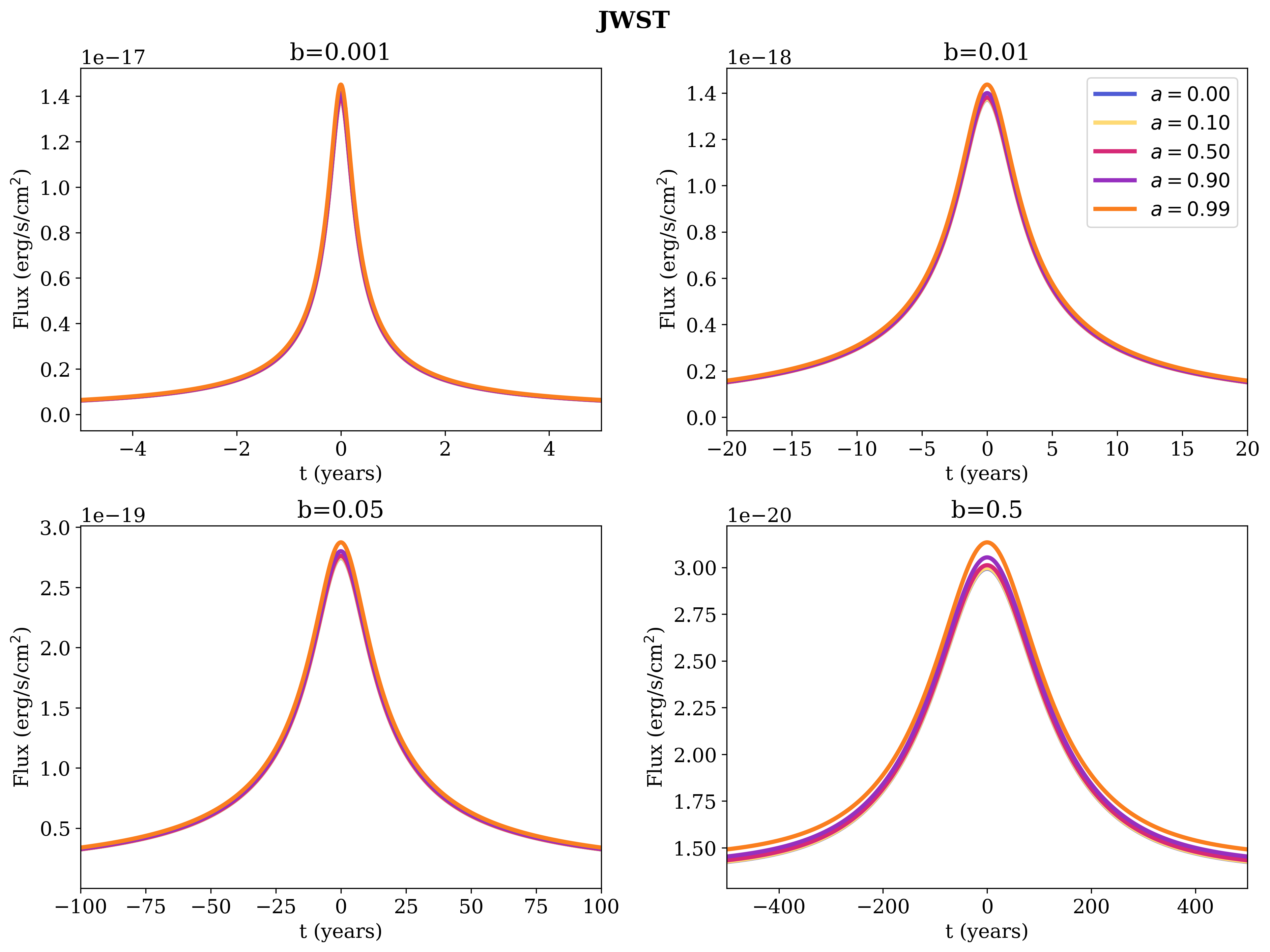}
    \caption{Simulated lensing events in the \textit{JWST} band, illustrating how impact parameter and spin  affect the magnification over time.}
    \label{fig:mag-jwst-spin}
\end{figure*}

\begin{figure*}
    \centering
    \includegraphics[width=\linewidth]{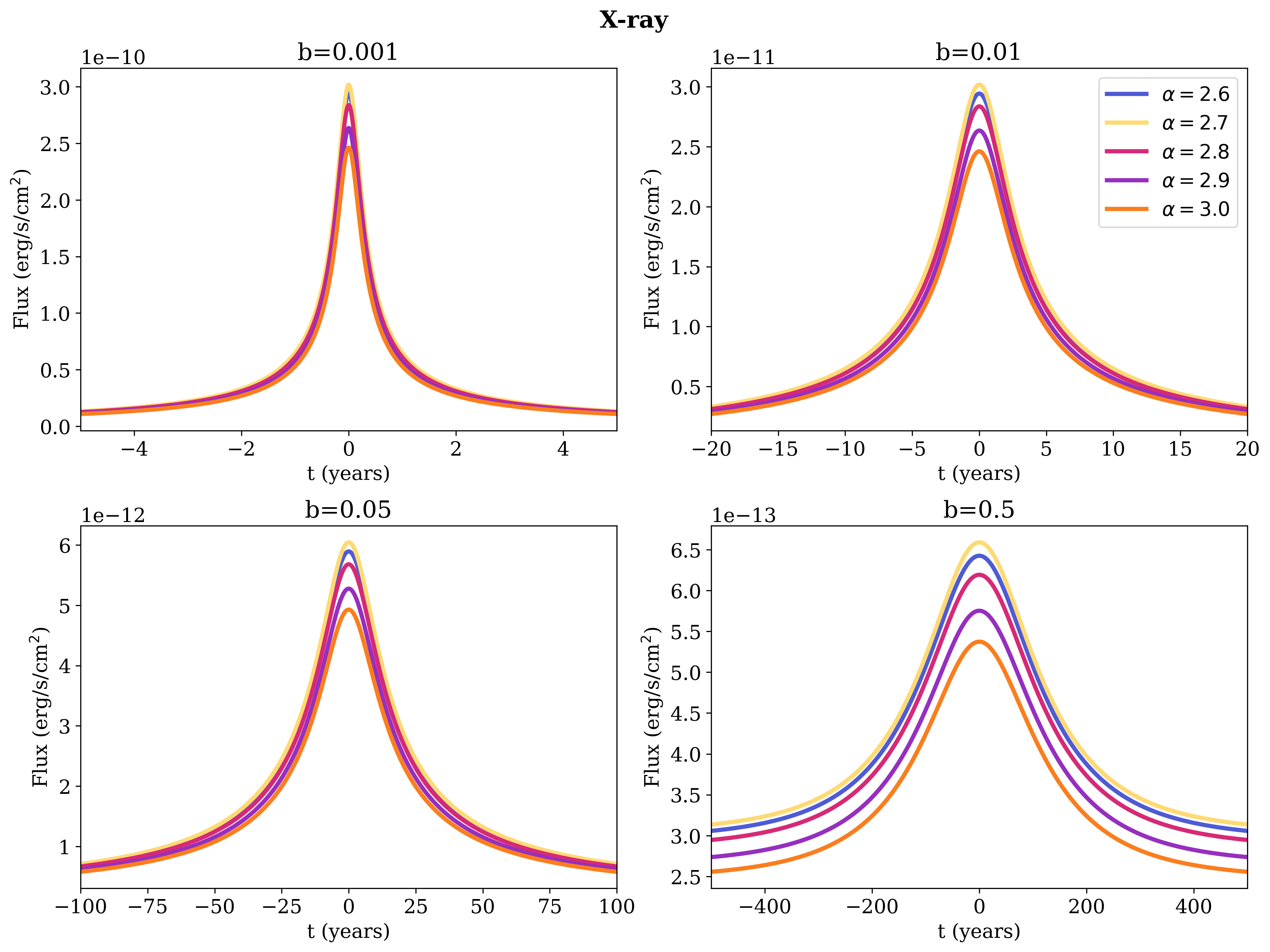}
    \caption{Simulated lensing events in the X-ray band, illustrating how impact parameter and temperature profile affect the magnification over time.}
    \label{fig:mag-xray-temp}
\end{figure*}

\begin{figure*}
    \centering
    \includegraphics[width=\linewidth]{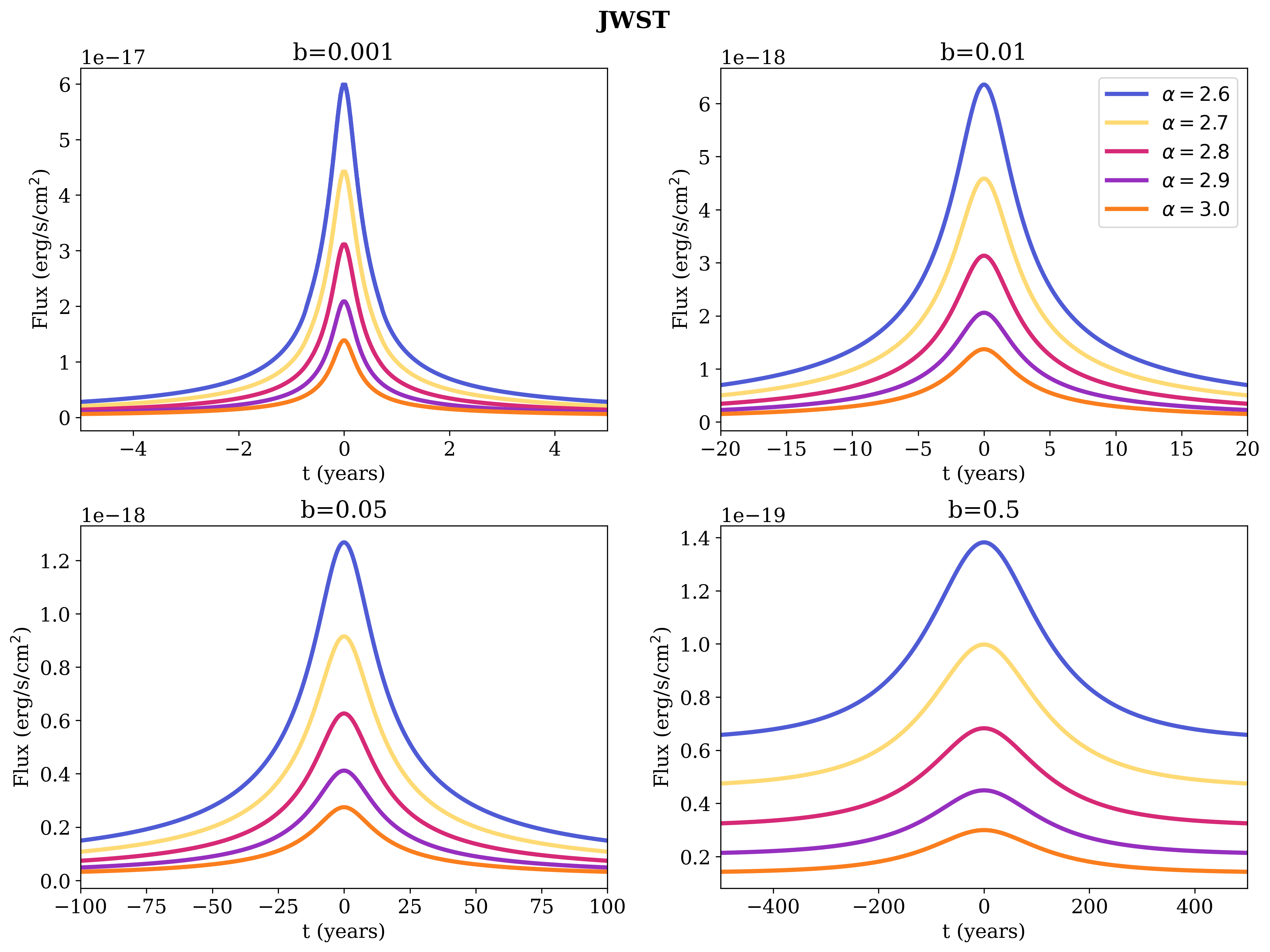}
    \caption{Simulated lensing events in the \textit{JWST} band, illustrating how impact parameter and temperature profile affect the magnification over time.}
    \label{fig:mag-jwst-temp}
\end{figure*}

One of the greatest challenges in modern astronomy is understanding how accretion operates, from the impact of spin, as determined via the location of the innermost stable circular orbit \citep[ISCO; see the reviews of][]{Middleton2016,Reynolds2021}, to the location of disc truncation at low-modest accretion rates \citep[e.g.][]{Conners2022}, to the impact of magnetic fields on the structure of the disc \citep[e.g][]{Fragile2023}, and even the radial temperature profile \citep[mostly studied in AGN, e.g.][]{Starkey2017,Neustadt2022}. Naturally, given the size scales involved, it is not possible to directly resolve accretion flows onto stellar mass black holes and neutron stars (even in our own galaxy) without deploying next generation, space-based interferometers (\citealt{uttley}). Lensing has already been used to infer the size of the X-ray corona in AGN (\citealt{chartas});. however, this involves a more complex system of lenses compared to microlensing.

To explore the potential power of lensing for resolving accretion flows in XRBs, we assume a standard temperature profile for a geometrically thin, optically thick disc: 

\begin{align}
    T_{\rm max} = f_{\rm col}\left\{ \frac{3GM\dot{M}}{8\pi R^\alpha\sigma_{SB}} \left[ 1-\left( \frac{R_{\rm isco}}{R} \right)^{1/2} \right]  \right\} ^{1/4} 
    \label{eq:temp}
\end{align}
\noindent where $M$ is the mass of the compact object, $\dot{M}$ is the accretion rate, $R$ is the radius in the disk at which the temperature is being evaluated, $R_{\rm isco}$ is the ISCO radius, $f_{\rm col}$ is the colour temperature correction set to a canonical value of 1.7 (\citealt{fcol}), and $\alpha$ determines the advection-dependent temperature profile of the disk, assumed to take a value between 2 and 3.





The radial flux profile within a given energy band is obtained by integrating a blackbody, with the peak temperature set by the above equation and flux scaled by 1/$f_{\rm col}^{4}$. 
We performed this numerically for a face-on accretion disk ($\theta = 0)$ using bands employed by the \textit{Zwicky Transient Facility} (\textit{ZTF}: $\lambda$ = 368-896 nm, \citealt{ztf}), \textit{JWST NIRCam} ($\lambda$ = 0.6-5 $\mu$m, \citealt{nircam}), {\it Swift UVOT} (170-650nm, \citealt{swift}) and \textit{XMM-Newton} ($E$ = 0.1-15~keV, \citealt{xmm}). We assume a canonical 10$M_\odot$ BH and an accretion rate of $10^{18}$ g/s (equivalent to an Eddington luminosity of $ L/L_{\rm Edd} \sim 10\%$).

In all cases, the upper integration limit, in units of $R_{\rm g}$ (= $GM/c^{2}$), is set at $10^6 R_{\rm g}$. We calculate the lower integration limit using the formulae for the innermost stable circular orbit (ISCO) around a rotating (Kerr) black hole:
\begin{align}
    Z_1 &= 1 + \left(1 - a^2\right)^{1/3} \left[ \left(1 + a\right)^{1/3} + \left(1 - a\right)^{1/3} \right] \\
    Z_2 &= \sqrt{2a^2 + Z_1^2} \\
    r_{\text{ISCO}} &= 3 + Z_2 \mp \sqrt{(3 - Z_1)(3 + Z_1 + 2Z_2)}
\end{align}
Here $a$, the dimensionless spin parameter, ranges from $0$ for a non-rotating (Schwarzschild) black hole to $1$ for a maximally rotating black hole. The sign of the last term depends on whether the orbit is prograde (minus) or retrograde (positive). 

In this paper, we ignore all relativistic effects except for gravitational redshift, which we account for  by shifting the upper and lower limits of the observed energy band by a factor of $( 1 - [ ( r R_g ) / ( r^2 + a^2 ) ] )^{1/2}$). This is the only significant relativistic effect for a face-on disc configuration. We will explore the signature of lensing for a range of disc inclinations with the full range of relativistic effects in a follow-up paper. 

The resulting flux maps are shown in Figure \ref{fig:heatmap}, labeled with the radius corresponding to the maximum flux in each band for a $10 M_{\odot}$ BH, for zero spin ($a = 0$), and for a temperature coefficient of $\alpha = 3$. As expected for an accretion disk, we observe the most intense emission in the X-ray regime within a compact region, whereas the emission detected by {\it Swift UVOT}, \textit{ZTF} and \textit{JWST} is from a larger, more diffuse region. 

Using these flux profiles, the time-dependent lensed flux profile within a given band can be found following the method of \cite{heyrovsky}, who used microlensing to measure the limb darkening of a lensed star. In brief, the magnification for each ring of thickness $dr$ at a distance $r$ from the center of the disk is found using the elliptic integrals $K(k)$ and $\Pi(n,k)$:

\begin{equation}
\begin{aligned}
A(l,r) &= \frac{4}{(l+r)\sqrt{(l-r)^2+4\epsilon^2}} \left[ 2\epsilon^2 K\left(\frac{4\epsilon}{l+r} \sqrt{\frac{lr}{(l-r)^2 + 4\epsilon^2}}\right) \right. \\
&\quad\left. +  \quad (l-r)^2 \Pi \left( \frac{4lr}{(l+r)^2} , \frac{4\epsilon}{l+r} \sqrt{\frac{lr}{(l-r)^2 + 4\epsilon^2}} \right) \right]
\end{aligned}
\end{equation}

\noindent where $l$ is the impact parameter divided by the maximum radius of the unlensed flux map ($r_{\rm max}$), and $\epsilon$ is the Einstein radius divided by $r_{\rm max}$. When the quantity $|\delta| = r - l$ is much less than 1, the following expression for the magnification is used:

\begin{equation}
\begin{aligned}
    A(l, l+\delta) \simeq \frac{2\epsilon}{l} (1 - \frac{\delta}{2l}) ln \frac{8\epsilon l}{|\delta| \sqrt{l^2 + \epsilon^2}} & + 4 \arctan \frac{l}{\epsilon} \\
    & + \frac{\epsilon(2l^2 + \epsilon^2}{l^2 (l^2+\epsilon^2)} \delta 
\end{aligned}
\end{equation}

For each position along the lens' trajectory, the lensed flux for one ring is found as the product of the flux of that ring (assuming symmetry) and the average $A(l,r)$ from all the points around the ring. The flux is then summed across all rings to find the total magnified flux from the whole disk at each lens position within the  band of interest. To allow the structure to be fully probed, we perform these calculations at a high resolution of $0.01 R_{\rm g}$ near the inner disk which then increases by 1\% with each step of our calculation, giving fine sampling close to the black hole and progressively coarser bins at larger radii.

Figures \ref{fig:mag-xray-spin} and \ref{fig:mag-jwst-spin} present the resulting lensing profiles for the accretion disk in the X-ray (0.3-10 keV) and infrared (\textit{JWST}) bands, assuming a WD velocity of 180 km/s, consistent with the velocity distribution shown in Figure \ref{fig:wd_dist}. We display the magnified fluxes for a range of impact parameters $b = [0.05, 0.5, 1.0, 2.0]$ and dimensionless spin parameters $a = [0.00, 0.10, 0.50, 0.90, 0.99]$. As expected, the strongest magnification occurs at the smallest impact parameter ($b = 0.05$) and highest spin ($a = 0.99$). For small impact parameters ($b < 1$), spin values above $a = 0.50$ lead to significant magnification. In contrast, for larger impact parameters ($b > 1$), the influence of spin on magnification becomes much less pronounced.

The X-ray results shown in Figure \ref{fig:mag-xray-spin} differ significantly in shape from the \textit{JWST} results in Figure \ref{fig:mag-jwst-spin}, indicating that observing an event in two bands enables detailed probing of the accretion disk’s  structure. Unsurprisingly, X-ray events are characterized by the sharpest peak in flux as the X-ray bright region is by far the most compact and produces very large magnifications when the lens crosses.

In addition to the spin, we also investigate how the temperature index, $\alpha$, affects the lensing profiles; this is shown in Figures \ref{fig:mag-xray-temp} and \ref{fig:mag-jwst-temp}. As before, we show these for impact parameters $b = [0.05, 0.5, 1.0, 2.0]$ and now for temperature index values of $\alpha = [2.6, 2.7, 2.8, 2.9, 3.0]$. We see that smaller values of $\alpha$ produce larger magnifications, and this effect is more pronounced for larger impact parameters. In Appendix \ref{sec:appendix} we show the same Figures for \textit{ZTF-r} and \textit{UV}.

\section{Discussion \& Conclusion}
\label{sec:discussion}

Lensing has proven to be a powerful technique in various fields but has not been deployed as a means by which to explore X-ray binaries. Based on an estimated number of WDs in the halo of M31, we have been able to predict the rate of X-ray lensing events, both within the {\it Swift} archive at the time of writing (2.6/year) and using the entire X-ray source population of M31 (6.3/year). 

Given the considerable probability that X-ray lensing has and will be observed towards M31, we have explored how such events might be used to access the details of the accretion flow. In this first study we have assumed a highly simplified geometry of a face-on thin accretion disc extending down to the ISCO and ignored relativistic effects except for gravitational redshift. In future we will extend our analysis to include other relativistic effects, inclination, disc truncation (both by the corona and by magnetic fields), other geometries (e.g. outflows) and irradiation. 

Our results suggest it may be feasible to recover both the black hole spin and accretion disk temperature profile by discerning their effects upon the profile of the microlensing magnification. As can be seen from Figures \ref{fig:mag-xray-spin} and \ref{fig:mag-jwst-spin} (spin) and Figures \ref{fig:mag-xray-temp} and \ref{fig:mag-jwst-temp} (temperature profile) these parameters have a significant effect upon the shape of the light curve. 
As expected, the high-energy X-ray emission (from the innermost disk) is most sensitive to black hole spin due to the spin-dependence of the ISCO. At smaller impact parameters, where the lens passes directly over the inner disk, the effect of spin on the magnification is most pronounced. Conversely, the lower-energy UV emission originates from the cooler, outer regions of the disk. Therefore, the temperature profile of the disk has a larger effect on the light curve at lower energies and for larger impact parameters. Observing a single microlensing event in multiple bands thereby allows us to fit both the high-energy and low-energy profiles simultaneously, providing a means to independently constrain both the black hole spin and the accretion disk temperature profile.

 A key question remains, however: is it possible to observe such events that would allow us to learn about such black hole properties? Our simulations are most sensitive to black hole spin and accretion disk temperature profile for events with very small impact parameters, specifically in the range of $b = 10^{-3} - 10^{-2}$. It is therefore crucial to determine if such highly aligned events are observable. While rare, these high-magnification events are not unprecedented. Many microlensing events observed in the MW have best fitting solutions for impact parameters of order $10^{-2}$ \citep[e.g.][]{yoo04,kruszynska24,chung25}, implying that it is indeed possible to observe the events with small impact parameters that we have simulated in this work. We also note the case of OGLE-2008-BLG-279, which reached a peak magnification of approximately 1600, corresponding to the smallest impact parameter we have used in this work of $b \approx 10^{-4}$ \citep{yee09}. With next-generation, high-cadence surveys such as the \textit{Galactic Bulge Time-Domain Survey} conducted by the \textit{Roman Space Telescope} \citep{schlieder24}, these rare, extreme microlensing events could allow us to glimpse the innermost regions of black hole accretion disks.

Our initial findings open a promising new direction for research in microlensing and the study of accretion flows in X-ray binaries. By generating a comprehensive suite of models across multiple wavelengths and systematically varying key parameters — such as lens mass, black hole spin, and impact parameter — it becomes possible to fit observational data and infer these physical properties, much like the approach used to determine parameters of stellar populations. This fitting can be performed using traditional MCMC techniques \citep[as done by e.g.][]{bagpipes,prospector} or through simulation-based inference, also known as likelihood-free inference \citep[e.g.][]{lovell19,patricia25}, aided by machine learning.

Although there is power in this technique, it is important to appreciate that not all X-ray sources in M31 are X-ray binaries and consider the likely impact this has on our predicted number of events for resolving the nature of accretion flows. Although not all X-ray sources in M31 have been identified with a specific type of system, multiple authors have classified a substantial proportion, with \cite{Lazzarini2018} finding 15 HMXB candidates (out of a total of 120 sources). 
\cite{Williams2018} detected 373 X-ray sources and identified optical counterpart candidates for 188 of these; extended background galaxies comprise more than half (107) of the counterpart candidates, 5 are star clusters, 12 are foreground stars, and 6 are supernova remnants. Since LMXBs are not as simple to identify due to faint optical counterparts, \cite{Williams2018} used stellar mass maps from \textit{PHAT} (\citealt{Williams2017}), to predict a LMXB population with $L_X > 3 \times 10^{35}$ erg/s of $\approx$ 100. Combining the results from \cite{Lazzarini2018} and \cite{Williams2018} with the total number of X-ray sources they detected, we estimate that the M31 X-ray population is composed of $12.5\%$ HMXBs, $26.8\%$ LMXBs, $28.7\%$ background galaxies, $3.2\%$ foreground stars, $1.6\%$ supernova remnants, and $27.2\%$ unclassified X-ray sources. We therefore take a conservative view and assume that only $\sim$ 40\% of known systems harbor accreting neutron stars and black holes. This implies a rate of lensing where there is an accretion disc present to $\sim$1/year in archival {\it Swift} data and $\sim$2.5/year when considering all X-ray sources in M31.

It is important to note that this first analysis assumes that the optical light from all systems is dominated by the accretion disc -- whilst this will not be strictly the case for HMXBs, we should still be able to explore the disc structure, as the lensed profiles should be separable 
 and distinct at different energies. We also note that the profiles of stellar and accretion disc lensing are very different, especially since accretion disk lensing will be observed in the X-ray band unlike stellar lensing. We have also assumed the X-ray sources to be persistent; this is obviously incorrect for the case of LMXBs which undergo periodic outbursts lasting for months to years (shorter variations such as type I bursts are unlikely to cause the same issues). During such outbursts, the disc geometry changes, with the longest time spent in the soft state \citep{done07}. This implies that lensing of HMXBs are most likely to probe the type of structure we have assumed but this will be explored in more detail in forthcoming papers.

\section*{Acknowledgements}
MM acknowledges support from STFC Consolidated grant (ST/V001000/1). The authors acknowledge the use of the IRIDIS High Performance Computing Facility, and associated support services at the University of Southampton. SN expresses gratitude to Mitchell Gerrard for engaging in insightful discussions regarding computational techniques.

\section*{Code and Data Availability}

The computational codes and data sets employed for generating the results presented in this paper are publicly accessible via the following online repository: \\ \texttt{\href{https://github.com/newmansoph/microlensing-in-M31}{github.com/sophie-newman/microlensing-in-M31}}

Additionally, the {\it Swift} and Chandra data used in this work can be found with the Xamin Web Interface on
NASA’s HEASARC (High Energy Astrophysics Science
Archive Research Center) website\footnote{\texttt{\href{https://heasarc.gsfc.nasa.gov/}{https://heasarc.gsfc.nasa.gov/}}}.



\bibliographystyle{mnras}
\bibliography{bib}

\begin{thebibliography}{}
\makeatletter
\relax
\def\mn@urlcharsother{\let\do\@makeother \do\$\do\&\do\#\do\^\do\_\do\%\do\~}
\def\mn@doi{\begingroup\mn@urlcharsother \@ifnextchar [ {\mn@doi@}
  {\mn@doi@[]}}
\def\mn@doi@[#1]#2{\def\@tempa{#1}\ifx\@tempa\@empty \href
  {http://dx.doi.org/#2} {doi:#2}\else \href {http://dx.doi.org/#2} {#1}\fi
  \endgroup}
\def\mn@eprint#1#2{\mn@eprint@#1:#2::\@nil}
\def\mn@eprint@arXiv#1{\href {http://arxiv.org/abs/#1} {{\tt arXiv:#1}}}
\def\mn@eprint@dblp#1{\href {http://dblp.uni-trier.de/rec/bibtex/#1.xml}
  {dblp:#1}}
\def\mn@eprint@#1:#2:#3:#4\@nil{\def\@tempa {#1}\def\@tempb {#2}\def\@tempc
  {#3}\ifx \@tempc \@empty \let \@tempc \@tempb \let \@tempb \@tempa \fi \ifx
  \@tempb \@empty \def\@tempb {arXiv}\fi \@ifundefined
  {mn@eprint@\@tempb}{\@tempb:\@tempc}{\expandafter \expandafter \csname
  mn@eprint@\@tempb\endcsname \expandafter{\@tempc}}}

\bibitem[\protect\citeauthoryear{{Alcock} et~al.,}{{Alcock}
  et~al.}{2000}]{alcock}
{Alcock} C.,  et~al., 2000, \mn@doi [\apj] {10.1086/309512}, \href
  {https://ui.adsabs.harvard.edu/abs/2000ApJ...542..281A} {542, 281}

\bibitem[\protect\citeauthoryear{{Ballard}, {Enzi}, {Collett}, {Turner}  \&
  {Smith}}{{Ballard} et~al.}{2024}]{ballard}
{Ballard} D.~J.,  {Enzi} W. J.~R.,  {Collett} T.~E.,  {Turner} H.~C.,   {Smith}
  R.~J.,  2024, \mn@doi [\mnras] {10.1093/mnras/stae514}, \href
  {https://ui.adsabs.harvard.edu/abs/2024MNRAS.528.7564B} {528, 7564}

\bibitem[\protect\citeauthoryear{{Beaulieu} et~al.,}{{Beaulieu}
  et~al.}{2006}]{exoplanets}
{Beaulieu} J.~P.,  et~al., 2006, \mn@doi [\nat] {10.1038/nature04441}, \href
  {https://ui.adsabs.harvard.edu/abs/2006Natur.439..437B} {439, 437}

\bibitem[\protect\citeauthoryear{{Belczynski}, {Kalogera}, {Rasio}, {Taam},
  {Zezas}, {Bulik}, {Maccarone}  \& {Ivanova}}{{Belczynski}
  et~al.}{2008}]{startrack}
{Belczynski} K.,  {Kalogera} V.,  {Rasio} F.~A.,  {Taam} R.~E.,  {Zezas} A.,
  {Bulik} T.,  {Maccarone} T.~J.,   {Ivanova} N.,  2008, \mn@doi [\apjs]
  {10.1086/521026}, \href
  {https://ui.adsabs.harvard.edu/abs/2008ApJS..174..223B} {174, 223}

\bibitem[\protect\citeauthoryear{Bellm et~al.,}{Bellm et~al.}{2018}]{ztf}
Bellm E.~C.,  et~al., 2018, \mn@doi [Publications of the Astronomical Society
  of the Pacific] {10.1088/1538-3873/aaecbe}, 131, 018002

\bibitem[\protect\citeauthoryear{{Birrer} et~al.,}{{Birrer}
  et~al.}{2020}]{holicow}
{Birrer} S.,  et~al., 2020, \mn@doi [\aap] {10.1051/0004-6361/202038861}, \href
  {https://ui.adsabs.harvard.edu/abs/2020A&A...643A.165B} {643, A165}

\bibitem[\protect\citeauthoryear{{Blaineau} et~al.,}{{Blaineau}
  et~al.}{2022}]{blaineau}
{Blaineau} T.,  et~al., 2022, \mn@doi [\aap] {10.1051/0004-6361/202243430},
  \href {https://ui.adsabs.harvard.edu/abs/2022A&A...664A.106B} {664, A106}

\bibitem[\protect\citeauthoryear{{Burrows} et~al.,}{{Burrows}
  et~al.}{2005}]{burrows}
{Burrows} D.~N.,  et~al., 2005, \mn@doi [\ssr] {10.1007/s11214-005-5097-2},
  \href {https://ui.adsabs.harvard.edu/abs/2005SSRv..120..165B} {120, 165}

\bibitem[\protect\citeauthoryear{{Calchi Novati}}{{Calchi
  Novati}}{2012}]{Calchi2012}
{Calchi Novati} S.,  2012, in Journal of Physics Conference Series. p. 012001
  (\mn@eprint {arXiv} {1201.5262}), \mn@doi{10.1088/1742-6596/354/1/012001}

\bibitem[\protect\citeauthoryear{{Carnall}, {McLure}, {Dunlop}  \&
  {Dav{\'e}}}{{Carnall} et~al.}{2018}]{bagpipes}
{Carnall} A.~C.,  {McLure} R.~J.,  {Dunlop} J.~S.,   {Dav{\'e}} R.,  2018,
  \mn@doi [\mnras] {10.1093/mnras/sty2169}, \href
  {https://ui.adsabs.harvard.edu/abs/2018MNRAS.480.4379C} {480, 4379}

\bibitem[\protect\citeauthoryear{{Chan} et~al.,}{{Chan} et~al.}{2020}]{chan}
{Chan} J. H.~H.,  et~al., 2020, \mn@doi [\aap] {10.1051/0004-6361/201937030},
  \href {https://ui.adsabs.harvard.edu/abs/2020A&A...636A..87C} {636, A87}

\bibitem[\protect\citeauthoryear{{Chartas}, {Kochanek}, {Dai}, {Poindexter}  \&
  {Garmire}}{{Chartas} et~al.}{2009}]{chartas}
{Chartas} G.,  {Kochanek} C.~S.,  {Dai} X.,  {Poindexter} S.,   {Garmire} G.,
  2009, \mn@doi [\apj] {10.1088/0004-637X/693/1/174}, \href
  {https://ui.adsabs.harvard.edu/abs/2009ApJ...693..174C} {693, 174}

\bibitem[\protect\citeauthoryear{{Chung} et~al.,}{{Chung}
  et~al.}{2025}]{chung25}
{Chung} S.-J.,  et~al., 2025, \mn@doi [arXiv e-prints]
  {10.48550/arXiv.2506.20914}, \href
  {https://ui.adsabs.harvard.edu/abs/2025arXiv250620914C} {p. arXiv:2506.20914}

\bibitem[\protect\citeauthoryear{{Collett} \& {Auger}}{{Collett} \&
  {Auger}}{2014}]{collett}
{Collett} T.~E.,  {Auger} M.~W.,  2014, \mn@doi [\mnras]
  {10.1093/mnras/stu1190}, \href
  {https://ui.adsabs.harvard.edu/abs/2014MNRAS.443..969C} {443, 969}

\bibitem[\protect\citeauthoryear{{Connors} et~al.,}{{Connors}
  et~al.}{2022}]{Conners2022}
{Connors} R. M.~T.,  et~al., 2022, \mn@doi [\apj] {10.3847/1538-4357/ac7ff2},
  \href {https://ui.adsabs.harvard.edu/abs/2022ApJ...935..118C} {935, 118}

\bibitem[\protect\citeauthoryear{{Crotts}}{{Crotts}}{1992}]{Crotts1992}
{Crotts} A. P.~S.,  1992, \mn@doi [\apjl] {10.1086/186602}, \href
  {https://ui.adsabs.harvard.edu/abs/1992ApJ...399L..43C} {399, L43}

\bibitem[\protect\citeauthoryear{{Dai}, {Chartas}, {Agol}, {Bautz}  \&
  {Garmire}}{{Dai} et~al.}{2003}]{Dai2003}
{Dai} X.,  {Chartas} G.,  {Agol} E.,  {Bautz} M.~W.,   {Garmire} G.~P.,  2003,
  \mn@doi [\apj] {10.1086/374548}, \href
  {https://ui.adsabs.harvard.edu/abs/2003ApJ...589..100D} {589, 100}

\bibitem[\protect\citeauthoryear{{Dai}, {Kochanek}, {Chartas}, {Koz{\l}owski},
  {Morgan}, {Garmire}  \& {Agol}}{{Dai} et~al.}{2010}]{Dai2010}
{Dai} X.,  {Kochanek} C.~S.,  {Chartas} G.,  {Koz{\l}owski} S.,  {Morgan}
  C.~W.,  {Garmire} G.,   {Agol} E.,  2010, \mn@doi [\apj]
  {10.1088/0004-637X/709/1/278}, \href
  {https://ui.adsabs.harvard.edu/abs/2010ApJ...709..278D} {709, 278}

\bibitem[\protect\citeauthoryear{{Done}, {Gierli{\'n}ski}  \& {Kubota}}{{Done}
  et~al.}{2007}]{done07}
{Done} C.,  {Gierli{\'n}ski} M.,   {Kubota} A.,  2007, \mn@doi [\aapr]
  {10.1007/s00159-007-0006-1}, \href
  {https://ui.adsabs.harvard.edu/abs/2007A&ARv..15....1D} {15, 1}

\bibitem[\protect\citeauthoryear{{Etherington} et~al.,}{{Etherington}
  et~al.}{2023}]{etherington}
{Etherington} A.,  et~al., 2023, \mn@doi [\mnras] {10.1093/mnras/stad582},
  \href {https://ui.adsabs.harvard.edu/abs/2023MNRAS.521.6005E} {521, 6005}

\bibitem[\protect\citeauthoryear{{Fantin} et~al.,}{{Fantin}
  et~al.}{2021}]{Fantin2021}
{Fantin} N.~J.,  et~al., 2021, \mn@doi [\apj] {10.3847/1538-4357/abf2b2}, \href
  {https://ui.adsabs.harvard.edu/abs/2021ApJ...913...30F} {913, 30}

\bibitem[\protect\citeauthoryear{{Fragile}, {Chatterjee}, {Ingram}  \&
  {Middleton}}{{Fragile} et~al.}{2023}]{Fragile2023}
{Fragile} P.~C.,  {Chatterjee} K.,  {Ingram} A.,   {Middleton} M.,  2023,
  \mn@doi [\mnras] {10.1093/mnrasl/slad099}, \href
  {https://ui.adsabs.harvard.edu/abs/2023MNRAS.525L..82F} {525, L82}

\bibitem[\protect\citeauthoryear{{Gaia Collaboration} et~al.,}{{Gaia
  Collaboration} et~al.}{2023}]{dr3}
{Gaia Collaboration} et~al., 2023, \mn@doi [\aap]
  {10.1051/0004-6361/202243940}, \href
  {https://ui.adsabs.harvard.edu/abs/2023A&A...674A...1G} {674, A1}

\bibitem[\protect\citeauthoryear{García-Berro, Torres, Isern  \&
  Burkert}{García-Berro et~al.}{1999}]{garcia-berro}
García-Berro E.,  Torres S.,  Isern J.,   Burkert A.,  1999, \mn@doi [Monthly
  Notices of the Royal Astronomical Society]
  {10.1046/j.1365-8711.1999.02115.x}, 302, 173

\bibitem[\protect\citeauthoryear{{Gilman}, {Du}, {Benson}, {Birrer},
  {Nierenberg}  \& {Treu}}{{Gilman} et~al.}{2020}]{gilman}
{Gilman} D.,  {Du} X.,  {Benson} A.,  {Birrer} S.,  {Nierenberg} A.,   {Treu}
  T.,  2020, \mn@doi [\mnras] {10.1093/mnrasl/slz173}, \href
  {https://ui.adsabs.harvard.edu/abs/2020MNRAS.492L..12G} {492, L12}

\bibitem[\protect\citeauthoryear{{Gould}}{{Gould}}{1996}]{Gould1996}
{Gould} A.,  1996, \mn@doi [\apj] {10.1086/177861}, \href
  {https://ui.adsabs.harvard.edu/abs/1996ApJ...470..201G} {470, 201}

\bibitem[\protect\citeauthoryear{{Gould}}{{Gould}}{2000}]{Gould2000}
{Gould} A.,  2000, \mn@doi [\apj] {10.1086/317037}, \href
  {https://ui.adsabs.harvard.edu/abs/2000ApJ...542..785G} {542, 785}

\bibitem[\protect\citeauthoryear{{Han} et~al.,}{{Han} et~al.}{2024}]{han24}
{Han} C.,  et~al., 2024, \mn@doi [\aap] {10.1051/0004-6361/202451806}, \href
  {https://ui.adsabs.harvard.edu/abs/2024A&A...692A.221H} {692, A221}

\bibitem[\protect\citeauthoryear{{Heyrovsk{\'y}}}{{Heyrovsk{\'y}}}{2003}]{heyrovsky}
{Heyrovsk{\'y}} D.,  2003, \mn@doi [\apj] {10.1086/376787}, \href
  {https://ui.adsabs.harvard.edu/abs/2003ApJ...594..464H} {594, 464}

\bibitem[\protect\citeauthoryear{{Hogg}}{{Hogg}}{2024}]{hogg}
{Hogg} N.~B.,  2024, \mn@doi [\mnras] {10.1093/mnrasl/slae005}, \href
  {https://ui.adsabs.harvard.edu/abs/2024MNRAS.529L..95H} {529, L95}

\bibitem[\protect\citeauthoryear{{Iglesias-Navarro} et~al.,}{{Iglesias-Navarro}
  et~al.}{2025}]{patricia25}
{Iglesias-Navarro} P.,  et~al., 2025, \mn@doi [arXiv e-prints]
  {10.48550/arXiv.2506.04336}, \href
  {https://ui.adsabs.harvard.edu/abs/2025arXiv250604336I} {p. arXiv:2506.04336}

\bibitem[\protect\citeauthoryear{{Jansen} et~al.,}{{Jansen} et~al.}{2001}]{xmm}
{Jansen} F.,  et~al., 2001, \mn@doi [\aap] {10.1051/0004-6361:20000036}, \href
  {https://ui.adsabs.harvard.edu/abs/2001A&A...365L...1J} {365, L1}

\bibitem[\protect\citeauthoryear{{Johnson}, {Leja}, {Conroy}  \&
  {Speagle}}{{Johnson} et~al.}{2021}]{prospector}
{Johnson} B.~D.,  {Leja} J.,  {Conroy} C.,   {Speagle} J.~S.,  2021, \mn@doi
  [\apjs] {10.3847/1538-4365/abef67}, \href
  {https://ui.adsabs.harvard.edu/abs/2021ApJS..254...22J} {254, 22}

\bibitem[\protect\citeauthoryear{{Kochanek}, {Dai}, {Morgan}, {Morgan}  \&
  {Poindexter}}{{Kochanek} et~al.}{2007}]{Kochanek2007}
{Kochanek} C.~S.,  {Dai} X.,  {Morgan} C.,  {Morgan} N.,   {Poindexter}
  S.~Chartas G.,  2007, in {Babu} G.~J.,  {Feigelson} E.~D.,  eds,
  Astronomical Society of the Pacific Conference Series Vol. 371, Statistical
  Challenges in Modern Astronomy IV. p.~43 (\mn@eprint {arXiv}
  {astro-ph/0609112}), \mn@doi{10.48550/arXiv.astro-ph/0609112}

\bibitem[\protect\citeauthoryear{{Koopmans}}{{Koopmans}}{2005}]{koopmans}
{Koopmans} L.~V.~E.,  2005, \mn@doi [\mnras]
  {10.1111/j.1365-2966.2005.09523.x}, \href
  {https://ui.adsabs.harvard.edu/abs/2005MNRAS.363.1136K} {363, 1136}

\bibitem[\protect\citeauthoryear{{Kruszy{\'n}ska} et~al.,}{{Kruszy{\'n}ska}
  et~al.}{2024}]{kruszynska24}
{Kruszy{\'n}ska} K.,  et~al., 2024, \mn@doi [\aap]
  {10.1051/0004-6361/202449322}, \href
  {https://ui.adsabs.harvard.edu/abs/2024A&A...692A..28K} {692, A28}

\bibitem[\protect\citeauthoryear{{La Barbera}, {Vazdekis}, {Ferreras}  \&
  {Pasquali}}{{La Barbera} et~al.}{2021}]{IMF}
{La Barbera} F.,  {Vazdekis} A.,  {Ferreras} I.,   {Pasquali} A.,  2021,
  \mn@doi [\mnras] {10.1093/mnras/stab1136}, \href
  {https://ui.adsabs.harvard.edu/abs/2021MNRAS.505..415L} {505, 415}

\bibitem[\protect\citeauthoryear{{Lazzarini} et~al.,}{{Lazzarini}
  et~al.}{2018}]{Lazzarini2018}
{Lazzarini} M.,  et~al., 2018, \mn@doi [\apj] {10.3847/1538-4357/aacb2a}, \href
  {https://ui.adsabs.harvard.edu/abs/2018ApJ...862...28L} {862, 28}

\bibitem[\protect\citeauthoryear{{Leauthaud} et~al.,}{{Leauthaud}
  et~al.}{2012}]{leauthaund}
{Leauthaud} A.,  et~al., 2012, \mn@doi [\apj] {10.1088/0004-637X/744/2/159},
  \href {https://ui.adsabs.harvard.edu/abs/2012ApJ...744..159L} {744, 159}

\bibitem[\protect\citeauthoryear{{Li}, {Collett}, {Krawczyk}  \& {Enzi}}{{Li}
  et~al.}{2024}]{tian}
{Li} T.,  {Collett} T.~E.,  {Krawczyk} C.~M.,   {Enzi} W.,  2024, \mn@doi
  [\mnras] {10.1093/mnras/stad3514}, \href
  {https://ui.adsabs.harvard.edu/abs/2024MNRAS.527.5311L} {527, 5311}

\bibitem[\protect\citeauthoryear{Licquia \& Newman}{Licquia \&
  Newman}{2015}]{mwmass}
Licquia T.~C.,  Newman J.~A.,  2015, \mn@doi [The Astrophysical Journal]
  {10.1088/0004-637X/806/1/96}, 806, 96

\bibitem[\protect\citeauthoryear{{Lovell}, {Acquaviva}, {Thomas}, {Iyer},
  {Gawiser}  \& {Wilkins}}{{Lovell} et~al.}{2019}]{lovell19}
{Lovell} C.~C.,  {Acquaviva} V.,  {Thomas} P.~A.,  {Iyer} K.~G.,  {Gawiser} E.,
    {Wilkins} S.~M.,  2019, \mn@doi [\mnras] {10.1093/mnras/stz2851}, \href
  {https://ui.adsabs.harvard.edu/abs/2019MNRAS.490.5503L} {490, 5503}

\bibitem[\protect\citeauthoryear{{Mandelbaum}, {Seljak}, {Kauffmann}, {Hirata}
  \& {Brinkmann}}{{Mandelbaum} et~al.}{2006}]{mandelbaum}
{Mandelbaum} R.,  {Seljak} U.,  {Kauffmann} G.,  {Hirata} C.~M.,   {Brinkmann}
  J.,  2006, \mn@doi [\mnras] {10.1111/j.1365-2966.2006.10156.x}, \href
  {https://ui.adsabs.harvard.edu/abs/2006MNRAS.368..715M} {368, 715}

\bibitem[\protect\citeauthoryear{Middleton}{Middleton}{2016}]{Middleton2016}
Middleton M.,  2016, Black Hole Spin: Theory and Observation.
Springer Berlin Heidelberg, Berlin, Heidelberg, pp 99--151,
  \mn@doi{10.1007/978-3-662-52859-4_3}

\bibitem[\protect\citeauthoryear{{Moniez}}{{Moniez}}{2001}]{nutshell}
{Moniez} M.,  2001, Cosmological physics with gravitational lensing:
  proceedings of the XXXVth Rencontres de Moriond, \href
  {https://ui.adsabs.harvard.edu/abs/2001cpgl.conf....3M} {pp 3--11}

\bibitem[\protect\citeauthoryear{{Morgan}, {Kochanek}, {Dai}, {Morgan}  \&
  {Falco}}{{Morgan} et~al.}{2008}]{Morgan2008}
{Morgan} C.~W.,  {Kochanek} C.~S.,  {Dai} X.,  {Morgan} N.~D.,   {Falco} E.~E.,
   2008, \mn@doi [\apj] {10.1086/592767}, \href
  {https://ui.adsabs.harvard.edu/abs/2008ApJ...689..755M} {689, 755}

\bibitem[\protect\citeauthoryear{{Morrison}}{{Morrison}}{1996}]{morrison}
{Morrison} H.~L.,  1996, in {Morrison} H.~L.,  {Sarajedini} A.,  eds,
  Astronomical Society of the Pacific Conference Series Vol. 92, Formation of
  the Galactic Halo...Inside and Out. p.~453

\bibitem[\protect\citeauthoryear{{Mr{\'o}z} et~al.,}{{Mr{\'o}z}
  et~al.}{2019}]{mroz19}
{Mr{\'o}z} P.,  et~al., 2019, \mn@doi [\apjs] {10.3847/1538-4365/ab426b}, \href
  {https://ui.adsabs.harvard.edu/abs/2019ApJS..244...29M} {244, 29}

\bibitem[\protect\citeauthoryear{{Nadler}, {Birrer}, {Gilman}, {Wechsler},
  {Du}, {Benson}, {Nierenberg}  \& {Treu}}{{Nadler} et~al.}{2021}]{nadler}
{Nadler} E.~O.,  {Birrer} S.,  {Gilman} D.,  {Wechsler} R.~H.,  {Du} X.,
  {Benson} A.,  {Nierenberg} A.~M.,   {Treu} T.,  2021, \mn@doi [\apj]
  {10.3847/1538-4357/abf9a3}, \href
  {https://ui.adsabs.harvard.edu/abs/2021ApJ...917....7N} {917, 7}

\bibitem[\protect\citeauthoryear{Neustadt \& Kochanek}{Neustadt \&
  Kochanek}{2022}]{Neustadt2022}
Neustadt J. M.~M.,  Kochanek C.~S.,  2022, \mn@doi [Monthly Notices of the
  Royal Astronomical Society] {10.1093/mnras/stac888}, 513, 1046

\bibitem[\protect\citeauthoryear{{Nunota} et~al.,}{{Nunota}
  et~al.}{2025}]{nunota24}
{Nunota} K.,  et~al., 2025, \mn@doi [\apj] {10.3847/1538-4357/ada352}, \href
  {https://ui.adsabs.harvard.edu/abs/2025ApJ...979..123N} {979, 123}

\bibitem[\protect\citeauthoryear{{Paulin-Henriksson}
  et~al.,}{{Paulin-Henriksson} et~al.}{2003}]{POINT-AGAPE2003}
{Paulin-Henriksson} S.,  et~al., 2003, \mn@doi [\aap]
  {10.1051/0004-6361:20030519}, \href
  {https://ui.adsabs.harvard.edu/abs/2003A&A...405...15P} {405, 15}

\bibitem[\protect\citeauthoryear{{Peng}, {Impey}, {Rix}, {Kochanek}, {Keeton},
  {Falco}, {Leh{\'a}r}  \& {McLeod}}{{Peng} et~al.}{2006}]{peng}
{Peng} C.~Y.,  {Impey} C.~D.,  {Rix} H.-W.,  {Kochanek} C.~S.,  {Keeton} C.~R.,
   {Falco} E.~E.,  {Leh{\'a}r} J.,   {McLeod} B.~A.,  2006, \mn@doi [\apj]
  {10.1086/506266}, \href
  {https://ui.adsabs.harvard.edu/abs/2006ApJ...649..616P} {649, 616}

\bibitem[\protect\citeauthoryear{{P{\'e}rigois}, {Belczynski}, {Bulik}  \&
  {Regimbau}}{{P{\'e}rigois} et~al.}{2021}]{2021PhRvD.103d3002P}
{P{\'e}rigois} C.,  {Belczynski} C.,  {Bulik} T.,   {Regimbau} T.,  2021,
  \mn@doi [\prd] {10.1103/PhysRevD.103.043002}, \href
  {https://ui.adsabs.harvard.edu/abs/2021PhRvD.103d3002P} {103, 043002}

\bibitem[\protect\citeauthoryear{{Poole} et~al.,}{{Poole} et~al.}{2008}]{swift}
{Poole} T.~S.,  et~al., 2008, \mn@doi [\mnras]
  {10.1111/j.1365-2966.2007.12563.x}, \href
  {https://ui.adsabs.harvard.edu/abs/2008MNRAS.383..627P} {383, 627}

\bibitem[\protect\citeauthoryear{{Pooley}, {Blackburne}, {Rappaport}  \&
  {Schechter}}{{Pooley} et~al.}{2007}]{Pooley2007}
{Pooley} D.,  {Blackburne} J.~A.,  {Rappaport} S.,   {Schechter} P.~L.,  2007,
  \mn@doi [\apj] {10.1086/512115}, \href
  {https://ui.adsabs.harvard.edu/abs/2007ApJ...661...19P} {661, 19}

\bibitem[\protect\citeauthoryear{{Reynolds}}{{Reynolds}}{2021}]{Reynolds2021}
{Reynolds} C.~S.,  2021, \mn@doi [\araa] {10.1146/annurev-astro-112420-035022},
  \href {https://ui.adsabs.harvard.edu/abs/2021ARA&A..59..117R} {59, 117}

\bibitem[\protect\citeauthoryear{{Rieke} et~al.,}{{Rieke}
  et~al.}{2023}]{nircam}
{Rieke} M.~J.,  et~al., 2023, \mn@doi [\pasp] {10.1088/1538-3873/acac53}, \href
  {https://ui.adsabs.harvard.edu/abs/2023PASP..135b8001R} {135, 028001}

\bibitem[\protect\citeauthoryear{{Ruiter}, {Belczynski}, {Benacquista}  \&
  {Holley-Bockelmann}}{{Ruiter} et~al.}{2007}]{ruiter}
{Ruiter} A.~J.,  {Belczynski} K.,  {Benacquista} M.,   {Holley-Bockelmann} K.,
  2007, \mn@doi [arXiv e-prints] {10.48550/arXiv.0712.0847}, \href
  {https://ui.adsabs.harvard.edu/abs/2007arXiv0712.0847R} {p. arXiv:0712.0847}

\bibitem[\protect\citeauthoryear{{Sahu} et~al.,}{{Sahu}
  et~al.}{2022}]{isolatedBH}
{Sahu} K.~C.,  et~al., 2022, \mn@doi [\apj] {10.3847/1538-4357/ac739e}, \href
  {https://ui.adsabs.harvard.edu/abs/2022ApJ...933...83S} {933, 83}

\bibitem[\protect\citeauthoryear{{Sahu} et~al.,}{{Sahu} et~al.}{2025}]{sahu25}
{Sahu} K.~C.,  et~al., 2025, \mn@doi [\apj] {10.3847/1538-4357/adbe6e}, \href
  {https://ui.adsabs.harvard.edu/abs/2025ApJ...983..104S} {983, 104}

\bibitem[\protect\citeauthoryear{{Sajadian}}{{Sajadian}}{2015}]{sunspots}
{Sajadian} S.,  2015, \mn@doi [\mnras] {10.1093/mnras/stv1349}, \href
  {https://ui.adsabs.harvard.edu/abs/2015MNRAS.452.2587S} {452, 2587}

\bibitem[\protect\citeauthoryear{{Schlieder} et~al.,}{{Schlieder}
  et~al.}{2024}]{schlieder24}
{Schlieder} J.~E.,  et~al., 2024, in {Coyle} L.~E.,  {Matsuura} S.,   {Perrin}
  M.~D.,  eds,  Society of Photo-Optical Instrumentation Engineers (SPIE)
  Conference Series Vol. 13092, Space Telescopes and Instrumentation 2024:
  Optical, Infrared, and Millimeter Wave. p. 130920S,
  \mn@doi{10.1117/12.3020622}

\bibitem[\protect\citeauthoryear{{Shajib} et~al.,}{{Shajib}
  et~al.}{2022}]{shajib}
{Shajib} A.~J.,  et~al., 2022, \mn@doi [arXiv e-prints]
  {10.48550/arXiv.2210.10790}, \href
  {https://ui.adsabs.harvard.edu/abs/2022arXiv221010790S} {p. arXiv:2210.10790}

\bibitem[\protect\citeauthoryear{{Shimura} \& {Takahara}}{{Shimura} \&
  {Takahara}}{1995}]{fcol}
{Shimura} T.,  {Takahara} F.,  1995, \mn@doi [\apj] {10.1086/175740}, \href
  {https://ui.adsabs.harvard.edu/abs/1995ApJ...445..780S} {445, 780}

\bibitem[\protect\citeauthoryear{{Sonnenfeld}, {Treu}, {Gavazzi}, {Suyu},
  {Marshall}, {Auger}  \& {Nipoti}}{{Sonnenfeld} et~al.}{2013}]{sonnenfeld}
{Sonnenfeld} A.,  {Treu} T.,  {Gavazzi} R.,  {Suyu} S.~H.,  {Marshall} P.~J.,
  {Auger} M.~W.,   {Nipoti} C.,  2013, \mn@doi [\apj]
  {10.1088/0004-637X/777/2/98}, \href
  {https://ui.adsabs.harvard.edu/abs/2013ApJ...777...98S} {777, 98}

\bibitem[\protect\citeauthoryear{{Starkey} et~al.,}{{Starkey}
  et~al.}{2017}]{Starkey2017}
{Starkey} D.,  et~al., 2017, \mn@doi [\apj] {10.3847/1538-4357/835/1/65}, \href
  {https://ui.adsabs.harvard.edu/abs/2017ApJ...835...65S} {835, 65}

\bibitem[\protect\citeauthoryear{{Sumi} et~al.,}{{Sumi} et~al.}{2013}]{sumi13}
{Sumi} T.,  et~al., 2013, \mn@doi [\apj] {10.1088/0004-637X/778/2/150}, \href
  {https://ui.adsabs.harvard.edu/abs/2013ApJ...778..150S} {778, 150}

\bibitem[\protect\citeauthoryear{{Torres}, {Cantero}, {Rebassa-Mansergas},
  {Skorobogatov}, {Jim{\'e}nez-Esteban}  \& {Solano}}{{Torres}
  et~al.}{2019}]{torres19}
{Torres} S.,  {Cantero} C.,  {Rebassa-Mansergas} A.,  {Skorobogatov} G.,
  {Jim{\'e}nez-Esteban} F.~M.,   {Solano} E.,  2019, \mn@doi [\mnras]
  {10.1093/mnras/stz814}, \href
  {https://ui.adsabs.harvard.edu/abs/2019MNRAS.485.5573T} {485, 5573}

\bibitem[\protect\citeauthoryear{{Torres}, {Rebassa-Mansergas}, {Camisassa}  \&
  {Raddi}}{{Torres} et~al.}{2021}]{torres21}
{Torres} S.,  {Rebassa-Mansergas} A.,  {Camisassa} M.~E.,   {Raddi} R.,  2021,
  \mn@doi [\mnras] {10.1093/mnras/stab079}, \href
  {https://ui.adsabs.harvard.edu/abs/2021MNRAS.502.1753T} {502, 1753}

\bibitem[\protect\citeauthoryear{{Udalski} et~al.,}{{Udalski}
  et~al.}{1994}]{udalski94}
{Udalski} A.,  et~al., 1994, \mn@doi [\actaa]
  {10.48550/arXiv.astro-ph/9407014}, \href
  {https://ui.adsabs.harvard.edu/abs/1994AcA....44..165U} {44, 165}

\bibitem[\protect\citeauthoryear{{Uttley} et~al.,}{{Uttley}
  et~al.}{2021}]{uttley}
{Uttley} P.,  et~al., 2021, \mn@doi [Experimental Astronomy]
  {10.1007/s10686-021-09724-w}, \href
  {https://ui.adsabs.harvard.edu/abs/2021ExA....51.1081U} {51, 1081}

\bibitem[\protect\citeauthoryear{{Vulic}, {Gallagher}  \& {Barmby}}{{Vulic}
  et~al.}{2016}]{catalogue}
{Vulic} N.,  {Gallagher} S.~C.,   {Barmby} P.,  2016, \mn@doi [\mnras]
  {10.1093/mnras/stw1523}, \href
  {https://ui.adsabs.harvard.edu/abs/2016MNRAS.461.3443V} {461, 3443}

\bibitem[\protect\citeauthoryear{{Watson} et~al.,}{{Watson}
  et~al.}{2009}]{watson09}
{Watson} M.~G.,  et~al., 2009, \mn@doi [\aap] {10.1051/0004-6361:200810534},
  \href {https://ui.adsabs.harvard.edu/abs/2009A&A...493..339W} {493, 339}

\bibitem[\protect\citeauthoryear{Wiktorowicz, Middleton, Khan, Ingram, Gandhi
  \& Dickinson}{Wiktorowicz et~al.}{2021}]{wikselflensing}
Wiktorowicz G.,  Middleton M.,  Khan N.,  Ingram A.,  Gandhi P.,   Dickinson
  H.,  2021, \mn@doi [Monthly Notices of the Royal Astronomical Society]
  {10.1093/mnras/stab2135}, 507, 374

\bibitem[\protect\citeauthoryear{{Williams} et~al.,}{{Williams}
  et~al.}{2017}]{Williams2017}
{Williams} B.~F.,  et~al., 2017, \mn@doi [\apj] {10.3847/1538-4357/aa862a},
  \href {https://ui.adsabs.harvard.edu/abs/2017ApJ...846..145W} {846, 145}

\bibitem[\protect\citeauthoryear{{Williams} et~al.,}{{Williams}
  et~al.}{2018}]{Williams2018}
{Williams} B.~F.,  et~al., 2018, \mn@doi [\apjs] {10.3847/1538-4365/aae37d},
  \href {https://ui.adsabs.harvard.edu/abs/2018ApJS..239...13W} {239, 13}

\bibitem[\protect\citeauthoryear{{Wyrzykowski} \& {Mandel}}{{Wyrzykowski} \&
  {Mandel}}{2020}]{Wyrzykowski2020}
{Wyrzykowski} {\L}.,  {Mandel} I.,  2020, \mn@doi [\aap]
  {10.1051/0004-6361/201935842}, \href
  {https://ui.adsabs.harvard.edu/abs/2020A&A...636A..20W} {636, A20}

\bibitem[\protect\citeauthoryear{{Wyrzykowski} et~al.,}{{Wyrzykowski}
  et~al.}{2023}]{2023Gaia}
{Wyrzykowski} {\L}.,  et~al., 2023, \mn@doi [\aap]
  {10.1051/0004-6361/202243756}, \href
  {https://ui.adsabs.harvard.edu/abs/2023A&A...674A..23W} {674, A23}

\bibitem[\protect\citeauthoryear{{Yee} et~al.,}{{Yee} et~al.}{2009}]{yee09}
{Yee} J.~C.,  et~al., 2009, \mn@doi [\apj] {10.1088/0004-637X/703/2/2082},
  \href {https://ui.adsabs.harvard.edu/abs/2009ApJ...703.2082Y} {703, 2082}

\bibitem[\protect\citeauthoryear{{Yoo} et~al.,}{{Yoo} et~al.}{2004}]{yoo04}
{Yoo} J.,  et~al., 2004, \mn@doi [\apj] {10.1086/381241}, \href
  {https://ui.adsabs.harvard.edu/abs/2004ApJ...603..139Y} {603, 139}

\bibitem[\protect\citeauthoryear{{Yuan}, {Zhu}, {Liu}, {Qu}  \& {Fan}}{{Yuan}
  et~al.}{2022}]{m31mass}
{Yuan} S.,  {Zhu} L.,  {Liu} C.,  {Qu} H.,   {Fan} Z.,  2022, \mn@doi [Research
  in Astronomy and Astrophysics] {10.1088/1674-4527/ac7af9}, \href
  {https://ui.adsabs.harvard.edu/abs/2022RAA....22h5023Y} {22, 085023}

\bibitem[\protect\citeauthoryear{{Zinn}}{{Zinn}}{1985}]{Zinn}
{Zinn} R.,  1985, \mn@doi [\apj] {10.1086/163249}, \href
  {https://ui.adsabs.harvard.edu/abs/1985ApJ...293..424Z} {293, 424}

\bibitem[\protect\citeauthoryear{{de Jong} et~al.,}{{de Jong}
  et~al.}{2006}]{MEGA2006}
{de Jong} J.~T.~A.,  et~al., 2006, \mn@doi [\aap] {10.1051/0004-6361:20053812},
  \href {https://ui.adsabs.harvard.edu/abs/2006A&A...446..855D} {446, 855}

\makeatother
\end{thebibliography}




\appendix

\section{Mock UVOT and ZTF events}
\label{sec:appendix}

\begin{figure*}
    \centering
    \includegraphics[width=\linewidth]{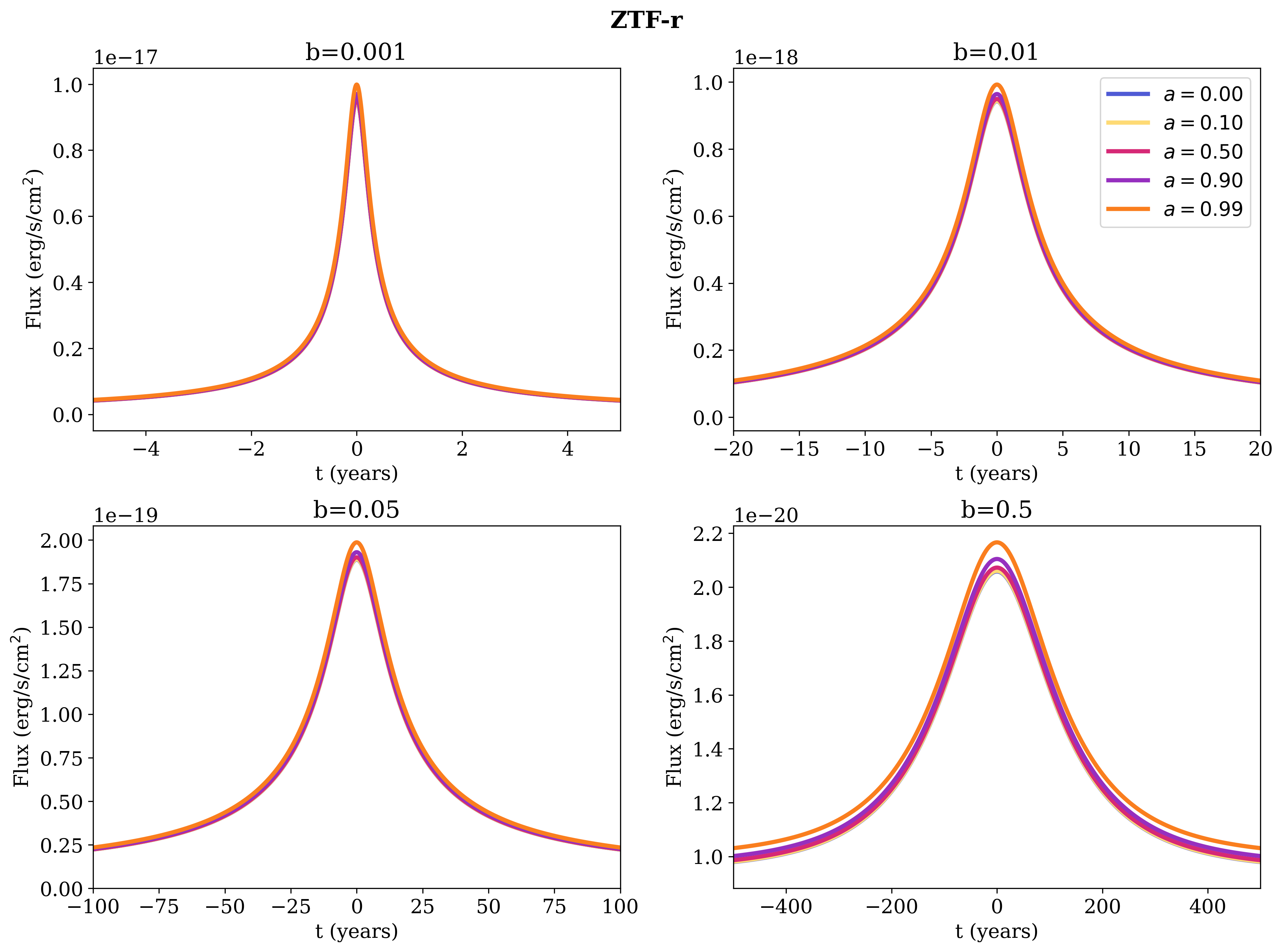}
    \caption{Simulated lensing events in the \textit{ZTF-r} band, illustrating how impact parameter and spin affect the magnification over time.}
    \label{fig:mag-ztf-spin}
\end{figure*}

\begin{figure*}
    \centering
    \includegraphics[width=\linewidth]{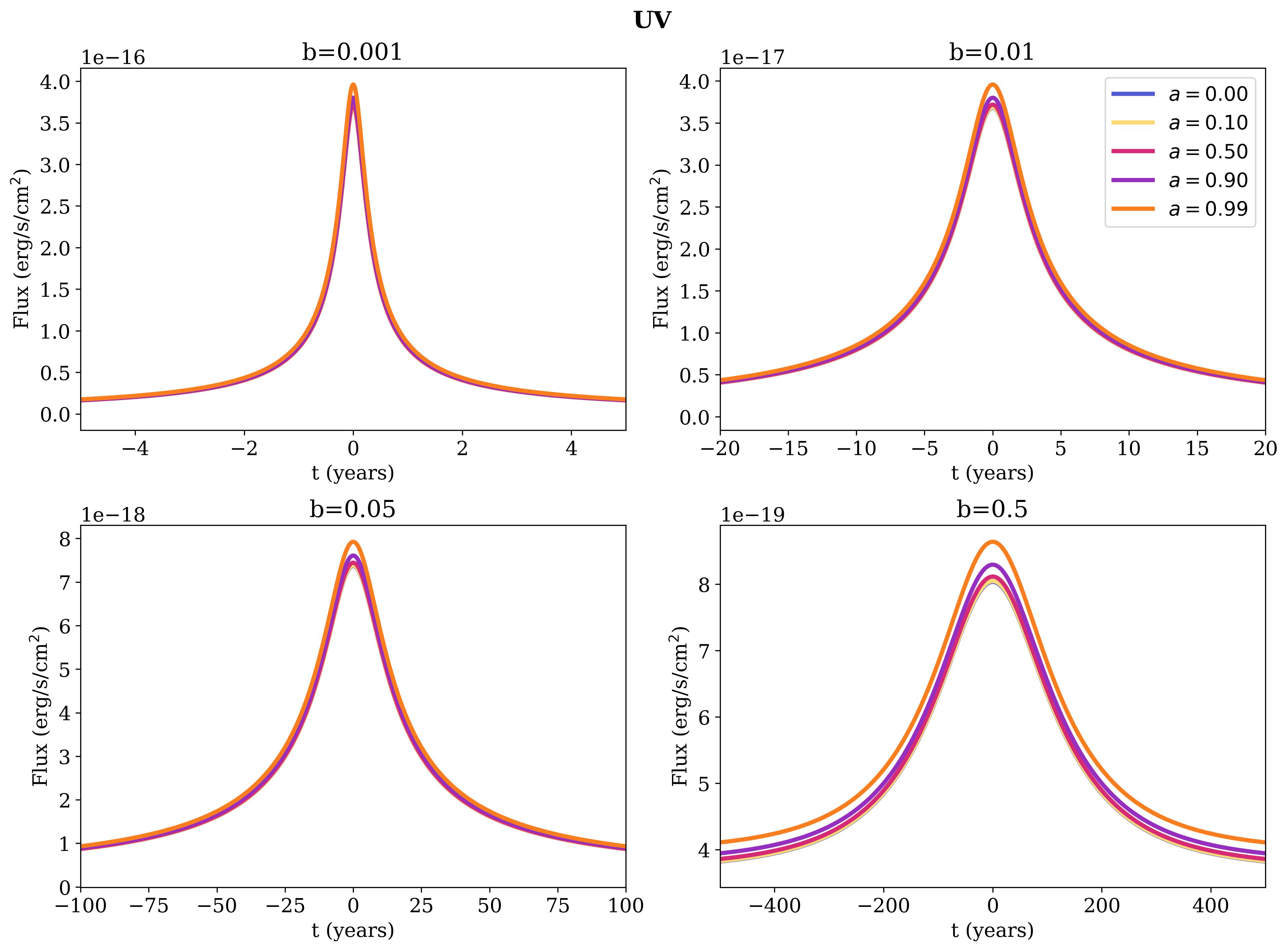}
    \caption{Simulated lensing events in the {\it Swift UVOT}  band, illustrating how impact parameter and spin affect the magnification over time.}
    \label{fig:mag-uv-spin}
\end{figure*}

\begin{figure*}
    \centering
    \includegraphics[width=\linewidth]{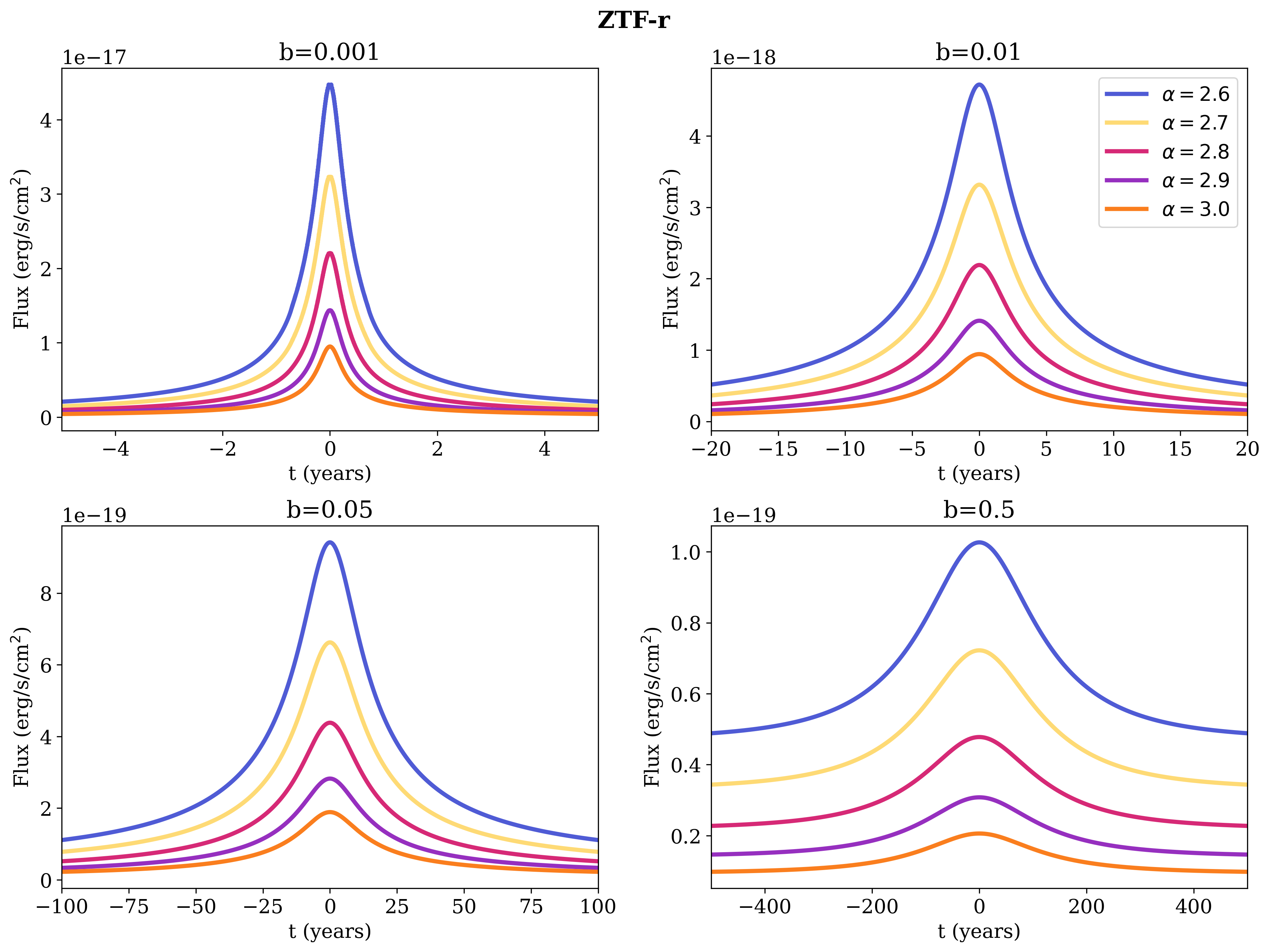}
    \caption{Simulated lensing events in the \textit{ZTF-r} band, illustrating how impact parameter and temperature profile affect the magnification over time.}
    \label{fig:mag-ztf-temp}
\end{figure*}

\begin{figure*}
    \centering
    \includegraphics[width=\linewidth]{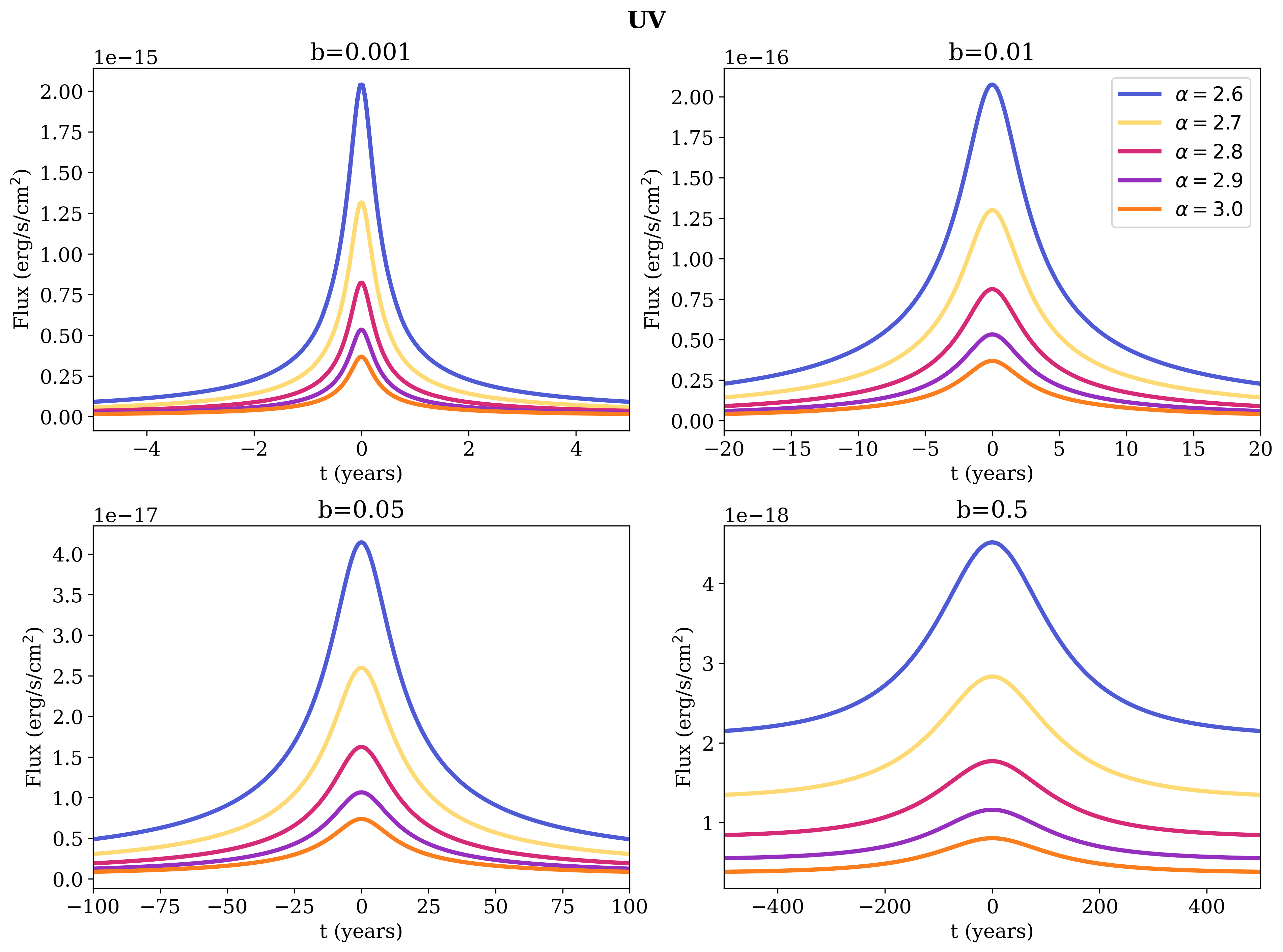}
    \caption{Simulated lensing events in the {\it Swift UVOT} band, illustrating how impact parameter and temperature profile affect the magnification over time.}
    \label{fig:mag-uv-temp}
\end{figure*}

Here we present the magnification profiles for the {\it Swift UVOT} and \textit{ZTF-r} bands. Figures \ref{fig:mag-ztf-spin} and \ref{fig:mag-uv-spin} show the effect of spin, while Figures \ref{fig:mag-ztf-temp} and \ref{fig:mag-uv-temp} show the effect of temperature.

For a fixed impact parameter of $b = 0.001$ and temperature profile of $\alpha = 3$, with a variable spin, we find that the greatest peak in flux is for the {\it XMM-Newton} light curve ($\sim 10^{-9}$ erg/s/cm$^2$), then \textit{UVOT} ($\sim 10^{-16}$ erg/s/cm$^2$), with \textit{JWST} ($\sim 5 \times 10^{-17}$ erg/s/cm$^2$) and \textit{ZTF-r} ($\sim 4 \times 10^{-17}$ erg/s/cm$^2$) with the smallest peak fluxes.

For a fixed impact parameter of $b = 0.001$ and spin of $a = 0$, and a variable temperature profile, we find that the greatest peak in flux is for the {\it XMM-Newton} light curve ($\sim 10^{-10}$ erg/s/cm$^2$), then {\it Swift UVOT} ($\sim 10^{-15}$ erg/s/cm$^2$), with \textit{JWST} ($\sim 5 \times 10^{-17}$ erg/s/cm$^2$) and \textit{ZTF-r} ($\sim 4 \times 10^{-17}$ erg/s/cm$^2$) with the smallest peak fluxes.

\bsp	
\label{lastpage}
\end{document}